\documentclass[12pt]{article}
\input epsf
\usepackage{epsfig,graphicx,subfigure}
\usepackage[dvips]{color}
\usepackage{amssymb}
\usepackage{bm}
\usepackage{amsmath}
\input epsf
\def\beqn{\begin{eqnarray}}
\def\eeqn{\end{eqnarray}}
\def\beq{\begin{equation}}
\def\eeq{\end{equation}}

\newcommand{\la}{\langle}
\newcommand{\ra}{\rangle}

\def\gevc2{(GeV/c)$^2$}

\newcommand{\be}{\begin{equation}}
\newcommand{\ee}{\end{equation}}
\newcommand{\ba}{\begin{eqnarray}}
\newcommand{\ea}{\end{eqnarray}}

\newcommand{\open}{{<\kern -0.3em{\scriptstyle )}}}

\begin{document}

\title{Probing Strangeness in Hard Processes
\\ \vspace{0.3cm} \small{\sf The science case of a RICH detector for CLAS12}}
\author{\small
H. Avakian$^{7}$,
M. Battaglieri$^{5}$,
E. Cisbani$^{6}$,
M. Contalbrigo$^{3}$,
U. D'Alesio$^{2}$,
R. De Leo$^{1}$,
R. Devita$^{5}$,
\\\small
P. Di Nezza$^{4}$,
D. Hasch$^{4}$,
V. Kubarovsky$^{7}$,
M. Mirazita$^{4}$,
M. Osipenko$^{5}$,
L. Pappalardo$^{3}$,
P. Rossi$^{4}$
%
%
%
\vspace*{0.5cm}
\\
\small{
$^{1}$Istituto Nazionale di Fisica Nucleare, Sezione di Bari and University of Bari, Bari, Italy}\\
\small{
$^{2}$Istituto Nazionale di Fisica Nucleare, Sezione di Cagliari and University of Cagliari, Cagliari, Italy}\\
\small{
$^{3}$Istituto Nazionale di Fisica Nucleare, Sezione di Ferrara, Ferrara, Italy}\\
\small{
$^{4}$Istituto Nazionale di Fisica Nucleare, Laboratori Nazionali di Frascati, Frascati, Italy}\\
\small{
$^{5}$Istituto Nazionale di Fisica Nucleare, Sezione di Genova, Genova, Italy}\\
\small{
$^{6}$Istituto Nazionale di Fisica Nucleare, Sezione di Roma, and Istituto Superiore di Sanita Roma, Italy}\\
\small{
$^{7}$Thomas Jefferson National Accelerator Facility, Newport News, VA, USA}
}
\date{\today}

\maketitle

\vspace*{1.5cm}
\centerline{Abstract}

\vspace*{0.3cm}

\small{
Since the discovery of strangeness almost five decades ago, interest in this degree of freedom 
has grown up and now its investigation spans the scales from quarks to nuclei. 
Measurements with identified strange hadrons can provide important information 
on several hot topics in hadronic physics: the strange distribution and fragmentation functions, 
the nucleon tomography and quark orbital momentum, accessible through the study of the 
{\it generalized} parton distribution and the {\it transverse momentum dependent} parton 
distribution functions, the quark hadronization in the nuclear medium, 
the hadron spectroscopy and the search for exotic mesons.\\
The CLAS12 large acceptance spectrometer in Hall B at the Jefferson Laboratory upgraded with a RICH detector 
together with the 12 GeV CEBAF high intensity, high polarized electron beam can open new 
possibilities to study strangeness in hard processes allowing breakthroughs in all those areas. 

This paper summarizes the physics case for a RICH detector for CLAS12. 
Many topics have been intensively discussed at the International Workshop 
"Probing Strangeness in Hard Processes" (PSHP2010)~\cite{PSHP-workshop}
held in Frascati, Italy in October 2010. 
The authors of this papers like to thank all speakers and participants of the workshop 
for their contribution and very fruitful discussion.
}
%
%


\clearpage
\tableofcontents
\clearpage


\section{Executive Summary}

\subsection {Nucleon structure and the role of strangeness}

Lepton scattering is the basic tool for determining the fundamental structure of matter,
in particu\-lar of the nucleon, from which the observable physical world around us
is formed.
Such experiments, using high energy electron beams, tested successfully the theory of 
Quantum Chromodynamics (QCD), which describes all strongly interacting matter in terms of
quark and gluon degrees of freedom.
The successful prediction of the energy dependence of parton distributions, which were introduced
to describe the structure of the nucleon, has been one of the great triumphs of pertubative QCD.

After four decades of lepton-nucleon scattering experiments, the gain in precision has often 
revealed intriguing aspects of the nucleon structure.
Among these surprises are the sizeable breaking of isospin symmetry in the light sea quark sector,
suggesting differences between the sea quark and antiquark distributions, the steep
rise of the distributions at small momentum fractions, and an interesting pattern of 
modifications of the distributions in nuclei.
Certainly one of the most surprising results is the unexpectedly small fraction, about a 
quarter, of the proton's spin that is due to the contribution from quarks and antiquarks.
This finding has triggered a vast experimental and theoretical activity aiming at clarifying
the role gluons and parton orbital angular momenta play for a complete description of the 
proton spin structure. 
New concepts of Transverse Momentum Dependent (TMD) distribution and fragmentation functions, 
which go beyond the collinear approximation, are a key to unravel the 
intricacies of the intrinsic motion of partons and the possible connection between their orbital
motion, their spin and the spin of the nucleon, which cannot be described with standard
(e.g. collinear) parton distributions.
These TMD distributions together with the so-called Generalized Parton Distributions (GPDs) provide
for the first time a framework to obtain information towards a genuine multi-dimensional momentum 
and space resolution of the nucleon structure.
This knowledge will have an important impact to other fields. 
Giving just one example, the information about the initial spatial distribution 
of quarks and gluons in the nucleon is essential for the interpretation of  heavy-ion collision data 
and the quest for the Quark-Gluon-Plasma.
The mapping of GPDs and TMDs and the deduction of a three-dimensional image of the 
nucleon is a major focus of the hadron physics community and constitutes a milestone in the physics 
program of the Jefferson Laboratory (JLab) 12 GeV upgrade\footnote{With the JLab 12 GeV upgrade,
up to 11 GeV electron beams will be delivered to Hall A, B and C and a 
12 GeV electron beam to Hall D.}~\cite{JLab12-TDR}.

While GPDs can be probed in hard exclusive processes observed in high energy lepton-nucleon scattering,
TMDs are most successfully measured in semi-inclusive deep-inelastic scattering (SIDIS).
In SIDIS experiments, also a hadron is detected in the final state in addition to the scattered lepton.
These experiments are the most powerful tool for directly obtaining flavour dependent information 
about the nucleon's quark structure. 
In particular, they provide unique access to the elusive strange distributions. 
Pioneering polarized semi-inclusive DIS experiments have revealed
surprising effects in various different kaon production observables, which deviate from
the expectations based on $u$-quark dominance for the scattering off a proton target.
These kaon results point to a significant role of sea quarks, and in particular strange quarks. 
For almost all kaon observables, the deviation from the expected behaviour is most pronounced in the 
kinematic region around $x_B=0.1$ ($x_B$ being the Bjorken scaling variable), 
which is well covered by CLAS12. 
In order to fully explore the power of SIDIS experiments, 
pion, kaon and proton separation over the full accessible kinematic range is essential.
\\
\vspace*{0.1cm}
\\
With the 12 GeV upgrade, JLab will provide the unique combination of 
high beam energy, high intensity (luminosity) and polarization, the usage of polarized targets
and advanced detection capabilities necessary for a mapping of the novel TMDs and 
GPDs. These functions will be uniquely explored in the valence kinematic region where many new, 
intriguing aspects of nucleon structure are expected to be most relevant. 
The addition of a RICH detector to the CLAS12 large acceptance spectrometer would make the upgraded 
Hall B an ideal place for carrying out these studies and
shading light on the elusive strange distribution and on the role sea quarks may
play in a complete, three-dimensional description of nucleon structure.

\subsection{Effects of the nuclear medium}
  
Besides the exciting new aspects of nucleon structure, 
a very interesting pattern of modifications of parton distribution and fragmentation functions
in nuclei has been observed, which caused vast experimental and theoretical activities.
The understanding of quark propagation in the nuclear medium is crucial for the interpretation
of high energy proton-nucleus interactions and ultrarelativistic heavy-ion collisions.
Leptoproduction of hadrons has the virtue that the energy
and momentum transferred to the hit parton are well determined, as it is 
``tagged'' by the scattered lepton and the nucleus is basically used as 
a probe at the fermi scale with increasing size or density, thus acting as 
femtometer-scale detectors of the hadronization process.
Theoretical models can therefore be calibrated in nuclear semi-inclusive DIS  and  
then applied, for example, to studies of the Quark-Gluon-Plasma.

The experimental results achieved over the last decade, demonstrate the enormous potential of 
nuclear SIDIS in shading light on the hadronization mechanisms.
For all observables investigated so far, a very distinct pattern of nuclear effects was observed 
for the various different hadron types.  
However, the existence and relative 
importance of the various stages, like the propagation and the interaction of the partons, 
color-neutralization and formation of the final hadron, are far from being determined unambiguously.

In this panorama, JLab12 with its high beam intensity and the usage of a large variety of nuclear 
targets will provide data in a kinematic region that is very suitable for studies
of nuclear effects. 
The potential of performing a fully differential analysis is a key to disentangle the
various different stages of hadronization.
The capability of identifying pions, kaons and protons over the whole kinematic range
of interest is essential for gaining more insights into the space-time evolution 
of the hadronization process.

\subsection{Search for exotic mesons}

The phenomenology of hadrons and in particular the study
of their spectrum led more than forty years ago to the
development of the quark model, where baryons and mesons
are described as bound systems of three quarks and of a
quark-antiquark pair, respectively.
Beyond these experimentally extensively observed states,
phenomenological models and lattice QCD
calculations suggest also the existence of exotic configurations such as
hybrids ($qqg$), tetraquarks ($qq\bar q\bar q$) and
glueballs.
The experimental verification of such exotic states would significantly deepen our
knowledge about the dynamics of QCD.
A very attractive method to identify exotic mesons is
through strangeness-rich final states,
where the kaons from the decay of the involved $\phi$-meson are usually high energetic.
Kaon identification over the whole accessible momentum range would
hence provide unique capabilities 
for the study of strangeonia and the search for exotic mesons.

\subsection{Impact of CLAS12 with a RICH}

The exciting physics program for CLAS12 is based on the unique features of the upgraded CEBAF and CLAS12 
spectrometer:

\begin{itemize}
\item high beam energy and intensity,
\item high beam polarization,
\item longitudinally and transversely polarized proton and effective neutron targets,
\item variety of nuclear targets,
\item large acceptance, multipurpose spectrometer.
\end{itemize}

\begin{figure}[t]
\begin{center}
\includegraphics[width=.58\textwidth]{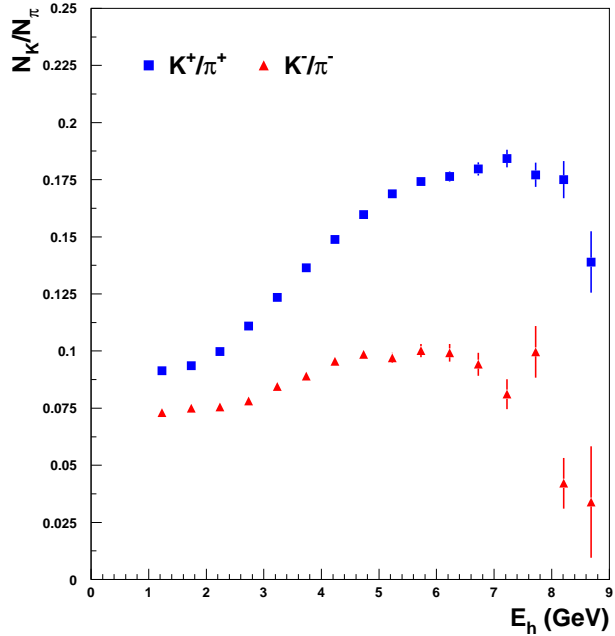}
\caption{\small 
The ratio of semi-inclusively produced kaons and pions as function of the kaon energy using
a Monte Carlo simulation with 11 GeV electron beam. The requirement on fractional hadron momentum $z_h >0.2$ 
is applied.  
}
\label{fig:pi-k-mom-distrib}
\end{center}
\end{figure}

\noindent
The addition of a RICH detector would significantly enhance the particle identification capabilities
of CLAS12 and make Hall B a unique place for studying the physics topics summarized above.
Figure~\ref{fig:pi-k-mom-distrib} presents the ratio of semi-inclusively produced kaons and pions as function
of the kaon energy using a Monte Carlo simulation with 11 GeV electron beam. 
The obtained distribution shows that kaon identification up to 8 GeV is higly desirable in 
order to fully explore the power of SIDIS experiments.
Furthermore, as pions greatly outnumber the other hadrons at nearly all kinematics, the RICH detector can 
tremendously reduce the backgrounds for the detection of unstable particles that decay to at 
least one charged non-pion. 

\clearpage

\section{\label{sec:longitudinal} The Longitudinal Structure of the Nucleon}
\subsection{Semi-inclusive DIS - a brief Introduction}

The exploration of the internal structure of hadrons in terms of 
quarks and gluons, the fundamental degrees of freedom of QCD, has been
and still is at the frontier of hadronic high energy physics.
Experiments
with inclusive Deep-Inelastic Scattering (DIS) processes, $\ell \, N
\to \ell' \, X$, have been performed for decades and have been
interpreted as the most common way to investigate the internal
structure of nucleons (protons and neutrons). Within the standard framework of
leading-twist perturbative QCD (pQCD) and its collinear
factorization theorems, the cross section for this inclusive
process, at large momentum transfer $Q^2$, can be expressed as a
convolution of parton distribution functions (PDF), $f_{a/p}(x)$,
giving the number density of partons, $a$, with a certain fraction $x$ of
the momentum of the parent hadron, $p$, with calculable elementary hard
interactions. 
The successful prediction of the scale ($Q^2$) dependence of PDFs has
been one of the great triumphs of pQCD.

A further and crucial insight in the internal structure of the nucleon
can be obtained by exploring its spin content. 
The  helicity distributions
 \begin{equation}
\Delta f_{a/N} = f_{+/+} - f_{-/+}\>,
 \end{equation}
where $f_{\pm/+}$ is the probability of finding a parton with
helicity $\pm$ in a nucleon with positive helicity,
can be studied in polarized DIS processes where both beam
lepton and target nucleon are longitudinally polarized.

Although very successful, inclusive DIS offers only limited
information about the internal nucleon structure and does not allow for a
direct flavour decomposition or an access to transverse degrees of freedom. 
A major and leading role in such
efforts is played by Semi-Inclusive Deep-Inelastic Scattering
(SIDIS) processes, $\ell \, N \to \ell' \, h \, X$, where in
addition to the scattered lepton, also a hadron is detected in the final state.
This hadron is generated in the fragmentation of the scattered quark
-- the so-called current fragmentation region. For such
processes the pQCD factorization theorem implies the following
expression:
\begin{equation}
  \frac{d\sigma^{\ell p \to \ell' h X}}{dx_Bdz_hdQ^2} = \sum_q f_{a/p}(x,Q^2) \otimes
d\hat\sigma^{\ell a \to \ell c} \otimes D_{h/c}(z_h,Q^2) \>,
\label{fac}
 \end{equation}
where
 \begin{equation}
 x_B = \frac {Q^2}{2p \cdot q}\quad\quad z_h = \frac{p
\cdot P_h}{p \cdot q} \quad\quad Q^2=-(l-l')^2 \> \cdot
 \end{equation}
Here, $l, l', p$ and $q$ are, respectively, the four-momenta of the incoming and 
scattered lepton, the target nucleon and the exchanged virtual boson.
In Eq.~(\ref{fac}) besides the PDF $f_{a/p}$, discussed above, a new
quantity, related to the hadronization of the scattered quark
appears: the fragmentation function (FF) $D_{h/c}(z_h)$, giving the
probability for a parton $c$ to produce a hadron $h$ with a fraction
$z_h$ of the momentum of the struck quark.

By studying a specific final hadron one can properly ``weight'' the
flavour of the incoming quark in the parent hadron and achieve a
more clear picture of the internal proton structure. Analogously,
polarized semi-inclusive DIS processes help for a much deeper exploration towards
a flavour decomposition of the longitudinal spin content of the
nucleon.

Both parton distribution and fragmentation functions cannot be calculated in pQCD which, 
however, predicts their scale dependence. 
This remarkable feature of pQCD together with the universality property of 
these functions, allows for their extraction from a suitable hard scattering process 
and their usage in phenomenological studies of any hard scattering process.
Famous examples are the calculation of inclusive hadron
production in hadron-hadron collisions, $pp\to h X$, or in $e^+e^-$
annihilation processes where only fragmentation functions enter the latter; or dilepton production
in hadron-hadron collision (Drell-Yan processes), which involves only distribution functions.
A more efficient strategy, still based
on the universality of distribution and fragmentation functions, is the approach of performing
global fits by combining simultaneously data from different
processes and at different scales.

While leading order
(LO) calculations simply extend the parton model expressions by
including proper pQCD scale dependence, 
higher order computations in the strong
coupling constant, and in particular next-to-leading order (NLO)
calculations, explored and validated for most processes, 
provide access to parton distributions for the various quark flavours and the gluon even from 
fully inclusive processes due to the different scaling behaviour of 
individual quarks and the gluon. 
NLO global analyses have become a very powerful tool for determining PDFs and FFs
provided the availabilty of data sets over a wide kinematic range
and, preferably, from various different processes.
In fact, the complementary information on the various partons obtained by combining 
measurements from different processes allows to relax and test rather stringent assumptions 
on the parameterizations adopted in the QCD fits and to probe the extracted functions
in various different energy regimes and kinematic configurations.

\subsection{Open issues for quark distributions at medium and high $x$}

Modern global analyses of parton distribution and fragmentation functions use data from a large variety 
of different processes in order to determine as much as possible different aspects of these 
functions (for a recent review see for example~\cite{Forte:2010dt}).
With the availability of HERA -- a QCD machine -- 
impressive progress was made over the last two decades in extracting parton distributions.
%
%

However, certain 
components even of the collinear structure of the nucleon are still poorly determined. 
Foremost among these is the elusive strangeness distribution.
But also the behaviour of parton distributions for $x \rightarrow 1$ is still under debate 
due to the lack of precise data in the very high $x$ region.

\subsubsection{Strangeness momentum distribution}

\begin{figure}[t]
\begin{center}
\includegraphics[width=.45\textwidth]{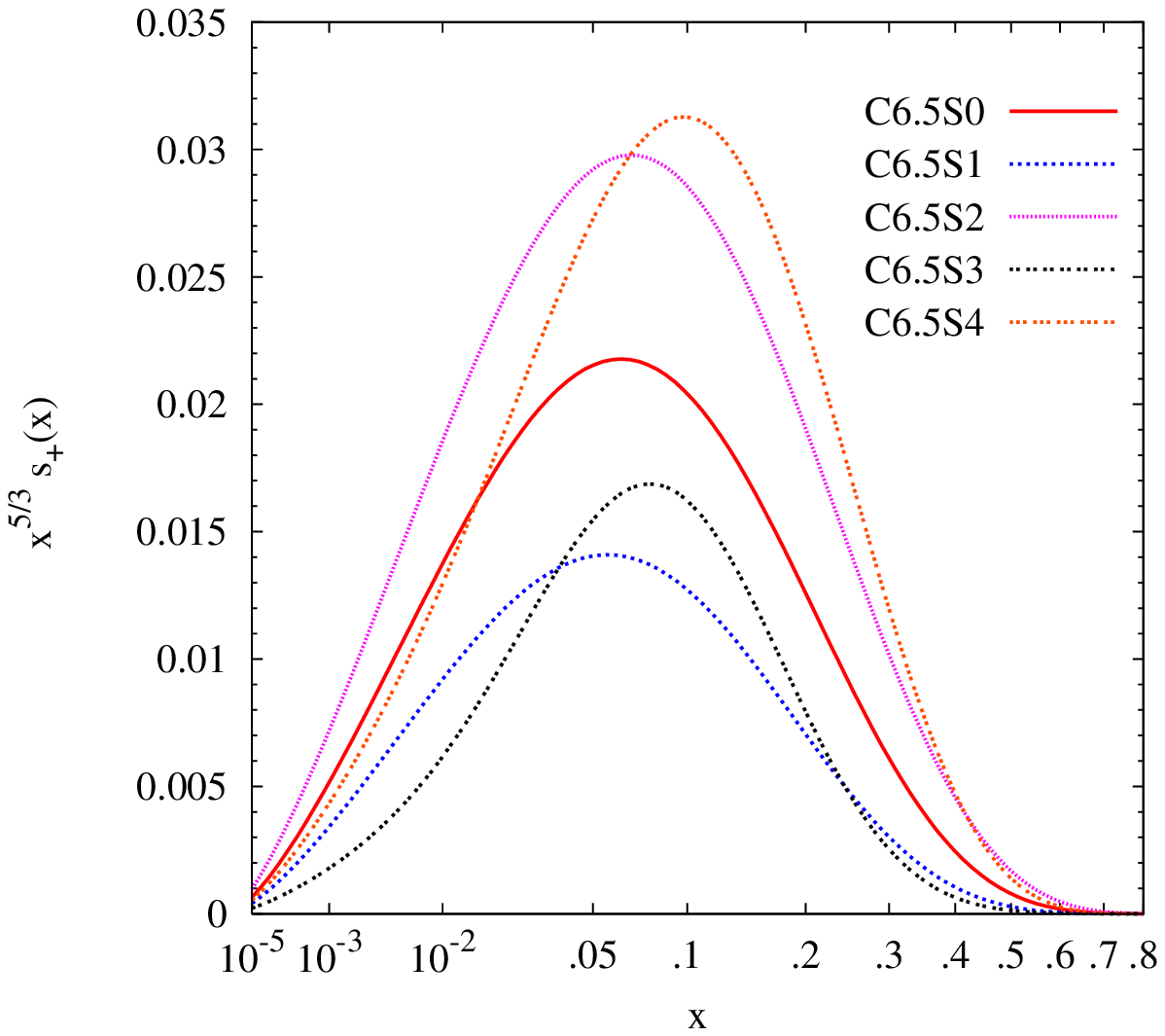}
\includegraphics[width=.45\textwidth]{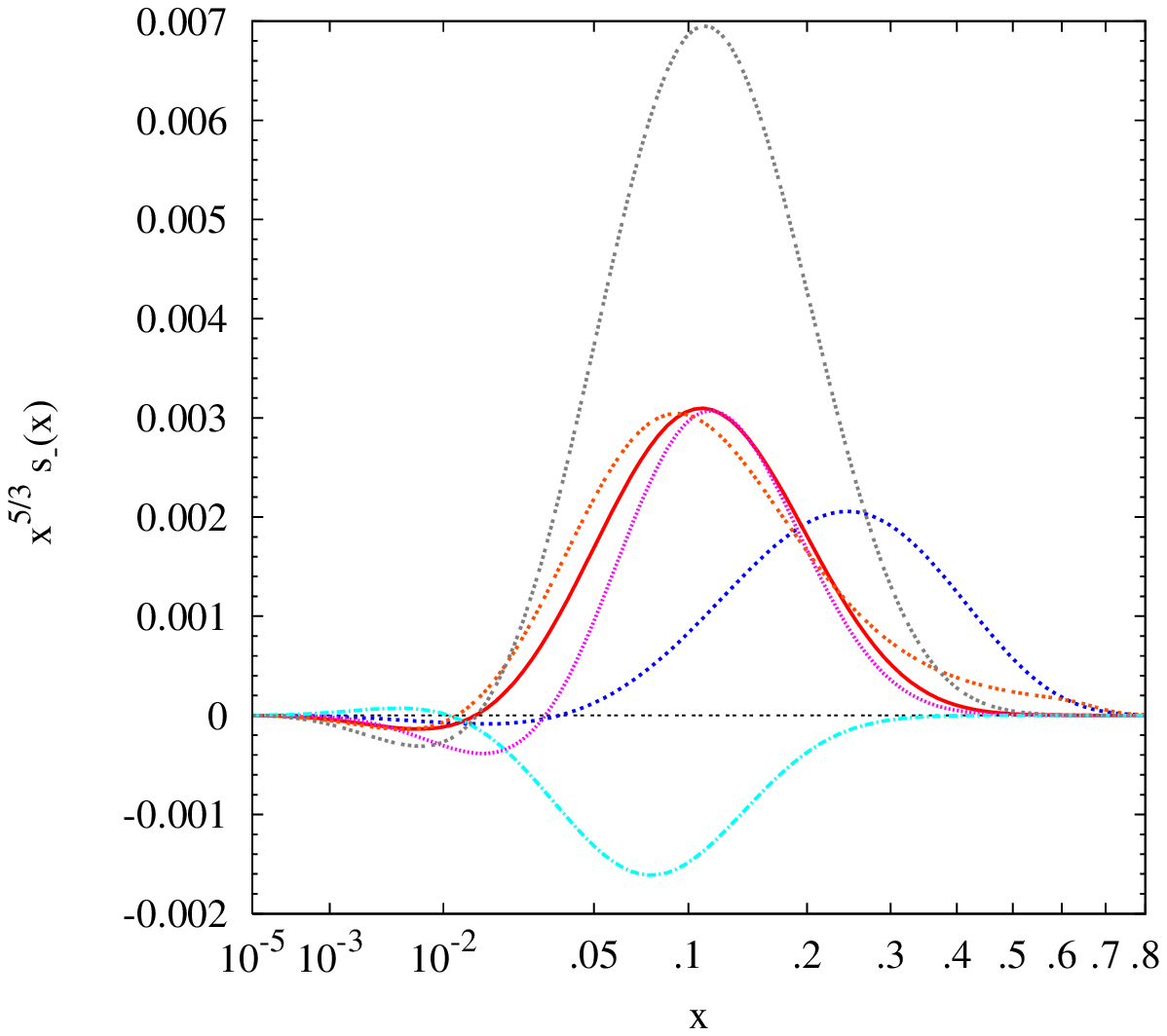}
\caption{\small 
Variations of the strange distributions $s^+=s+\bar s$ (left) and
$s^-=s-\bar s$ (right) for the five PDF sets CTEQ6.5Si, $i=0,\dots,4$~\cite{Lai:2007dq}
that are consistent with the data used in the fits.
}
\label{CTEQ-strange}
\end{center}
\end{figure}

\begin{figure}[t]
\begin{center}
\includegraphics[width=.49\textwidth]{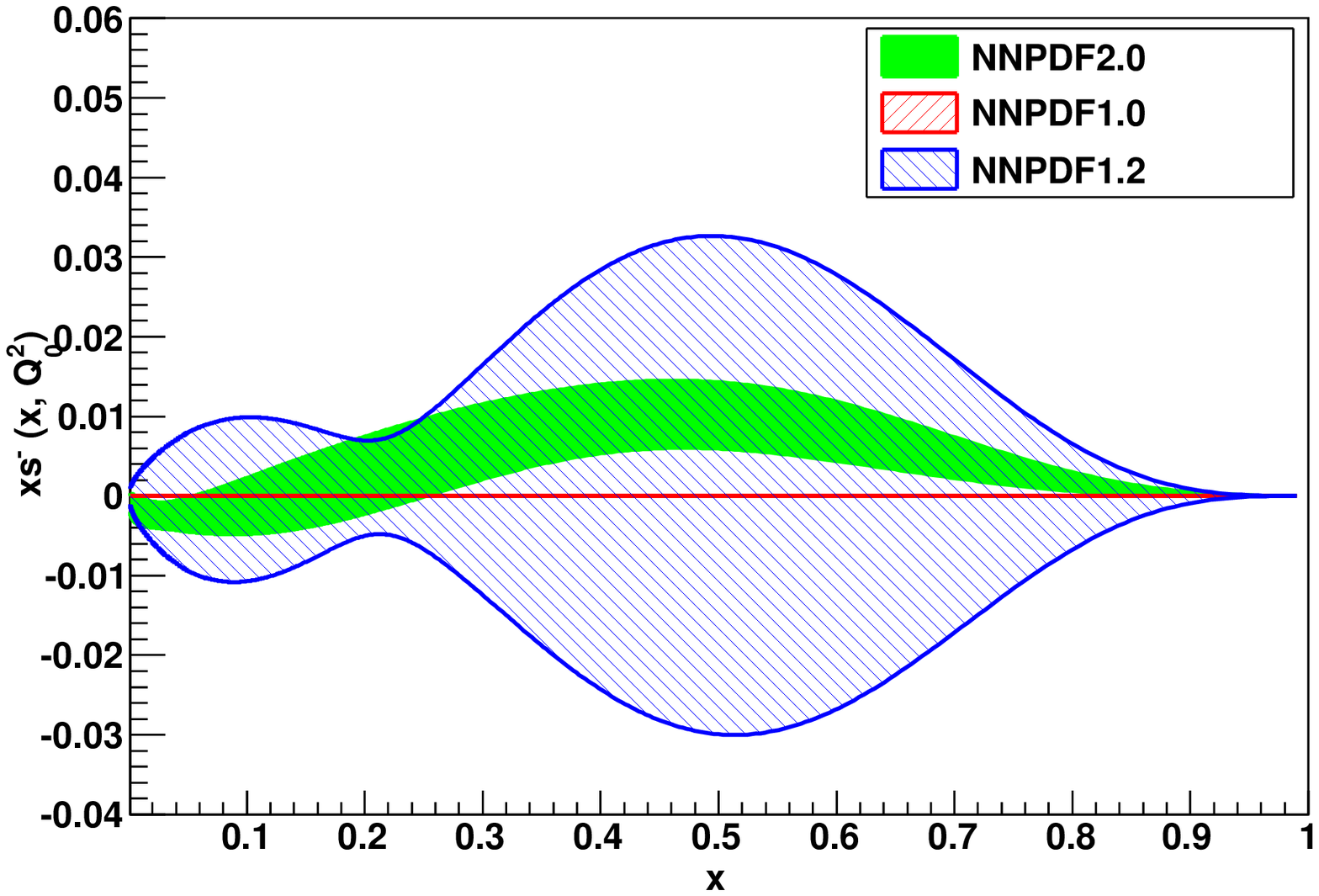}
\includegraphics[width=.49\textwidth]{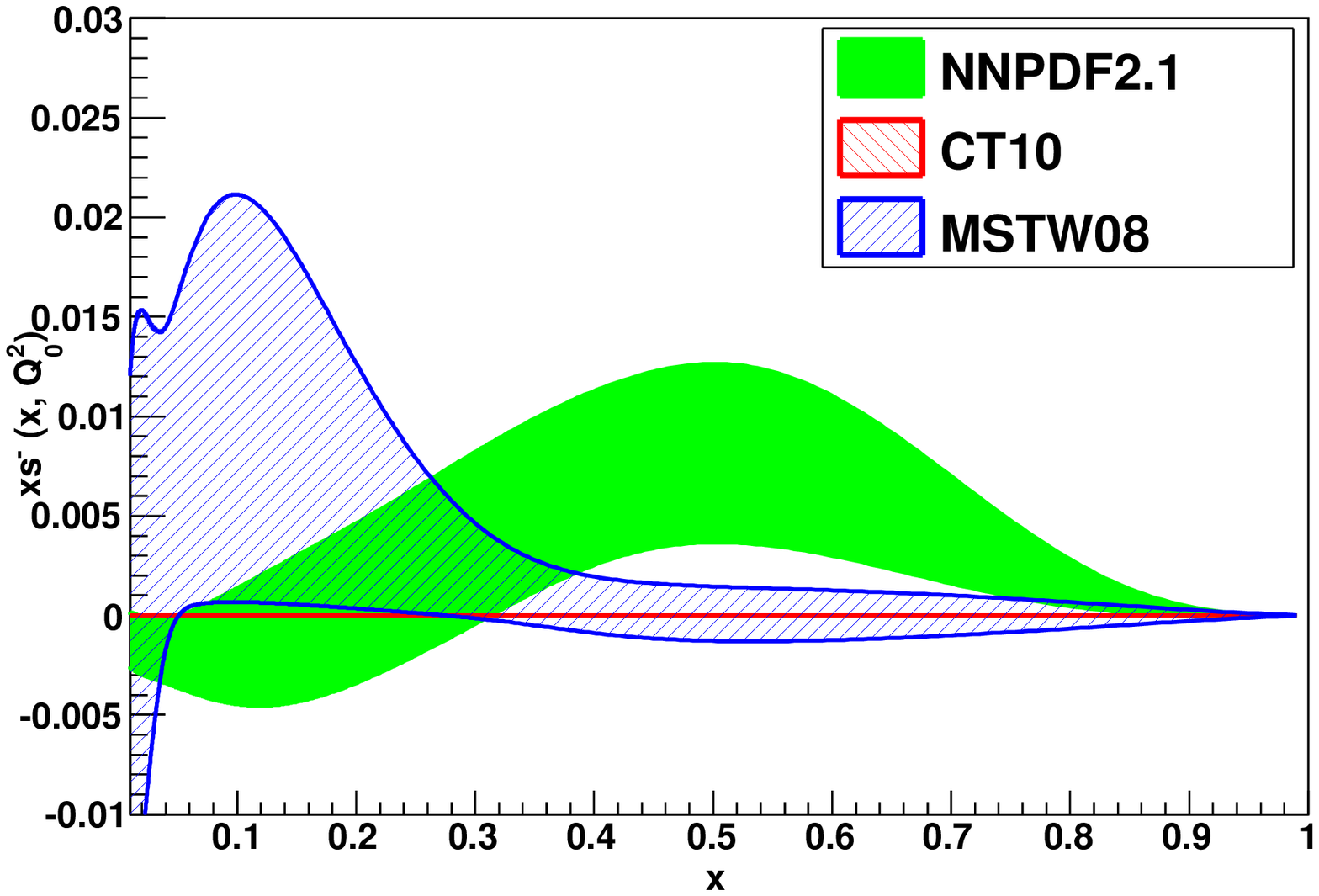}
\caption{\small
The strange distribution
$s^-=s-\bar s$. Left: the NNPDF extraction~\cite{Ball:2010de} of $xs^-$ determined in
  a fit to DIS including dimuon data (NNPDF2.0) and not including them 
  (NNPDF1.2). A fit without dimuon data where the strange distribution is
  fixed by the assumption $\bar s = s=C_S/2(\bar{u}+\bar{d})$ with $C_s=0.5$ 
  (NNPDF1.0) is also shown by the full (red) line.
Right: most recent NNPDF extraction of $xs^-$ (NNPDF2.1~\cite{Ball:2011uy}) compared with extractions
from the CTEQ (CT10~\cite{Lai:2010vv}) and MSTW (MSTW08~\cite{Martin:2009iq}) groups.} 
\label{NNPDF-strange}
\end{center}
\end{figure}

The determination of strangeness is challenging as it has the same electroweak couplings
as the down distribution while it is typically much smaller than it.
In absence of significant experimental constraints, the lack of knowledge is
reflected in the common practice of adopting the simplified ansatz 
$\bar s = s=C_s/2(\bar{u}+\bar{d})$, where even the proportionality constant $C_s$ is
only very loosely constrained by data.
The only way of determining the strange distribution accurately from data is to include 
semi-inclusive information.
Useful but limited information is provided by neutrino and antineutrino charm production 
(known as dimuon production) which is sensitive to strange distributions through the LO 
partonic process $W^+ + s \rightarrow c$.
The limitations in constraining the two strange combinations $s^+=s+\bar{s}$ and $s^-=s-\bar{s}$ 
are illustrated in Fig.~\ref{CTEQ-strange} and Fig.~\ref{NNPDF-strange} showing extractions
of the strange distributions
by the CTEQ-group and the NNPDF-group, respectively.
All shown extractions also include the information from the neutrino dimuon data.
In Fig.~\ref{CTEQ-strange}, the solid (red) curve shows the reference PDF set CTEQ6.5S0.
The other curves illustrate the range of variation of the magnitude and shape of $s^\pm(x)$
that are consistent with the data used in the fits~\cite{Lai:2007dq}. 

Figure~\ref{NNPDF-strange} (left) also shows the result for $s^-$ obtained when assuming the strange 
distribution being proportional to the light quark sea, i.e. $\bar s = s=C_s/2(\bar{u}+\bar{d})$. 
This result is then misleadingly accurate as demonstrated by the full (red) line (NNPDF1.0). 
\begin{figure}[hb]
\begin{center}
\includegraphics[width=.55\textwidth]{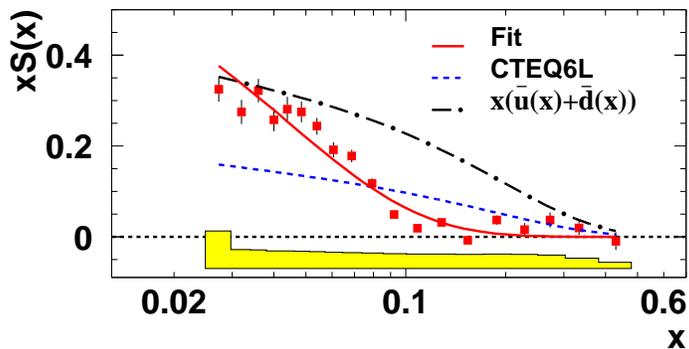}
\caption{\small 
The strange distributions $S=s^+=s+\bar s$ extracted in a LO analysis by the HERMES
experiment~\cite{Airapetian:2008qf} using data from SIDIS kaon production.
The solid curve is a 3-parameter fit for $S(x)=x^{a}e^{-x/b}(1-x)$ and
the dot-dashed curve is the sum of light antiquarks from CTEQ6L.
The shape of $S$ is found to be very different from the average of the light sea. 
}
\label{hermes-strange}
\end{center}
\end{figure}
Very recently, the NNPDF-group derived a new, by a factor of two in the uncertainty 
improved, strange distribution by comparing  Drell-Yan cross section data 
above and below the charm threshold~\cite{Ball:2010de} (labelled as NNPDF2.0 or NNPDF2.1 in 
Fig.~\ref{NNPDF-strange}).
The most recent NNPDF extraction of $s^-=s-\bar s$ (NNPDF2.1~\cite{Ball:2011uy}) is compared in
Fig.~\ref{NNPDF-strange} (right) with those from 
the CTEQ (CT10~\cite{Lai:2010vv}) and MSTW (MSTW08~\cite{Martin:2009iq}) groups.
Here, the NNPDF strangeness distribution is parameterized with as many parameters as any other
PDF resulting an uncertainty that is mostly larger than the one 
of MSTW08 and CT10 strangeness sets, which use very few parameters.
Clearly, much more data is needed for pinning down the strangeness distribution.

\vspace*{0.2cm}

Complementary and new information about the strange content of the nucleon is provided 
by kaon electroproduction data from semi-inclusive DIS.
The high potential of information contained in such data has been illustrated 
by the HERMES collaboration which performed a
LO extraction of the strange distribution $S=s^+=s+\bar s$ from charged-kaon 
production in DIS on the deuteron~\cite{Airapetian:2008qf}, 
shown in Fig.~\ref{hermes-strange}. 
This result confirms that the strange distribution is substantially different from 
the average of the light sea quarks.

Such data from charged kaon production in DIS were already successfully included in NLO
global QCD analyses of helicity distributions where they provide 
stringent constraints for a flavour separation, as discussed in the next section.
Their inclusion in global QCD analyses of momentum distributions is still an open 
but promising task, in particular in view of new SIDIS data
which will be available in near future from COMPASS and CLAS12 experiments.  
As seen in Fig.~\ref{hermes-strange}, the deviation of the strangness distribution from 
the expected behaviour is most pronounced in the kinematic range around $x=0.1$ which is well 
covered by CLAS12. 
Kaon identification over the whole momentum range will be an essential ingredient for shading light
on the elusive strange distribution.

\subsubsection{Strangeness helicity distribution}

\begin{figure}[t]
\begin{center}
\mbox{
\hspace*{-0.5cm}
\includegraphics[width=.5\textwidth]{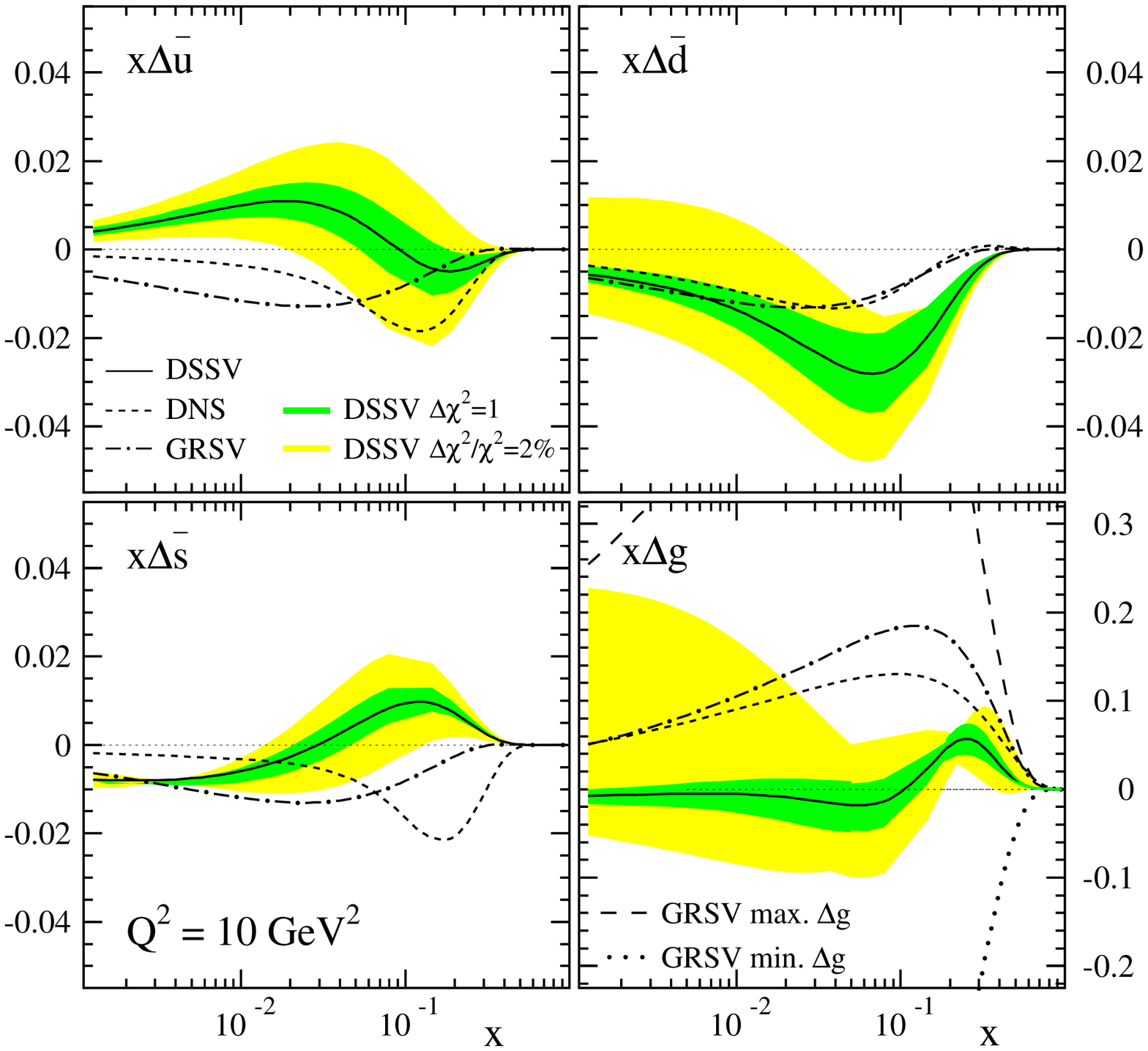}
\hspace*{-0.5cm}
\includegraphics[width=.55\textwidth]{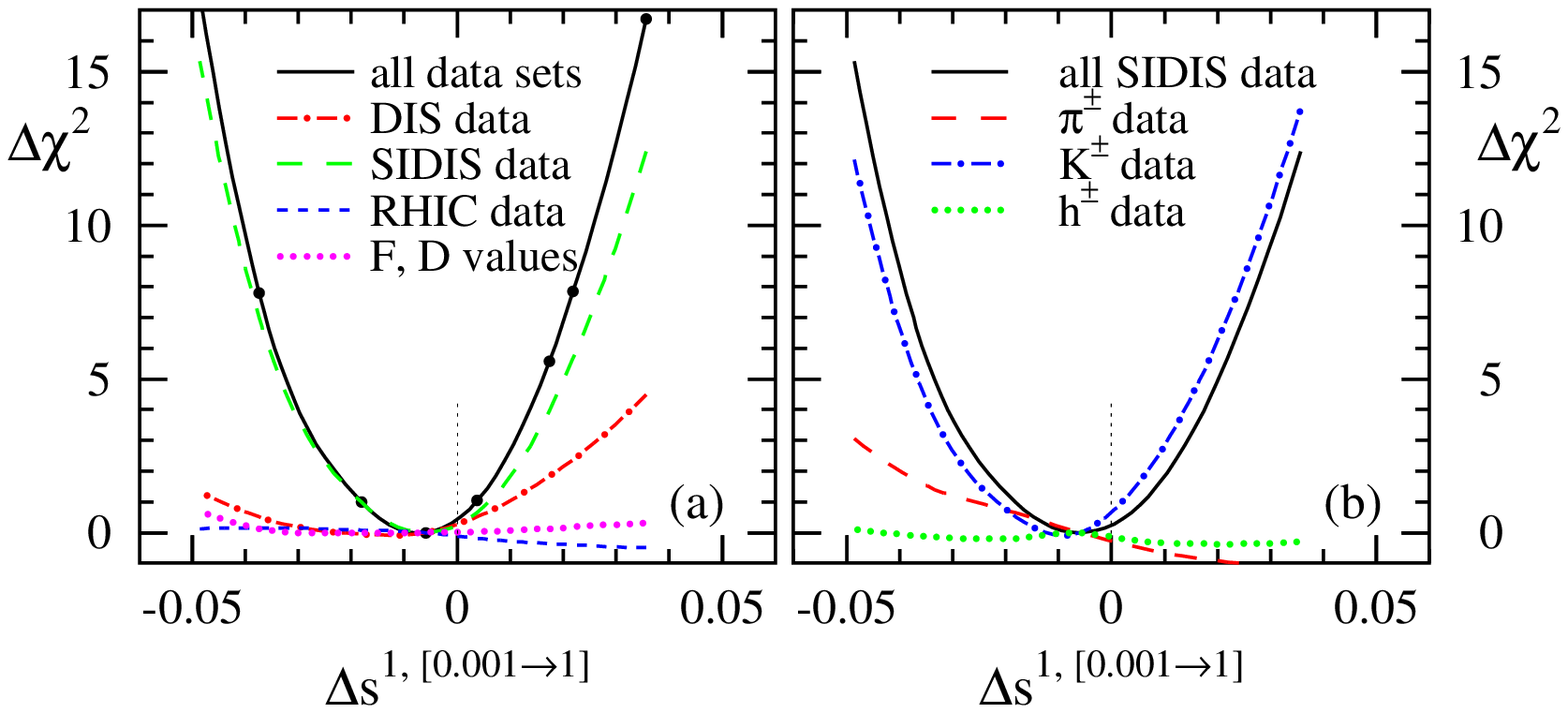}
}
\caption{\small 
Left: DSSV polarized sea and gluon densities compared with previous fits 
~\cite{Gluck:2000dy,deFlorian:2005mw,Navarro:2006bb}. 
The shaded bands correspond to alternative fits with 
uncertainties $\Delta\chi^2 = 1$ and $\Delta\chi^2/\chi^2 = 2\%$.
Right:
the $\chi^2$ profiles and partial contributions $\Delta \chi^2$  
of the various types of data sets for variations of the truncated first moment 
$\Delta s^{1}$ for the range $0.001 < x < 1.0$. 
}
\label{DSSV-full-set}
\end{center}
\end{figure}
Modern global QCD analyses of helicity parton distributions make use of data from
both inclusive and semi-inclusive polarized DIS, as well as from polarized
proton-proton ($pp$) scattering at RHIC.
In absence of inclusive {\it polarized} DIS data from a {\it collider}, sea quark and 
gluon helicity distributions can only be poorly constrained from scaling 
violation.
Therefore, from the early days of analyses of polarized data, there was an
attempt to use data that provide more direct flavour information and/or
access to the gluon polarization.
But only recently, a comprehensive, fully NLO QCD analysis of available polarized 
data, including $pp$ scattering data, has been developed by the 
DSSV-group~\cite{deFlorian:2008mr,deFlorian:2009vb}.
This analysis benefited significantly from the improved knowledge of parton-to-hadron
fragmentation functions derived by the DSS-group~\cite{deFlorian:2007aj}.
For the first time, these fragmentation functions provide a good description of identified 
hadron yields in the entire kinematic range relevant for the analysis of polarized semi-inclusive DIS and
$pp$ scattering data. 

While the up and down quark helicity distributions could be determined with rather good 
precision in the range $0.001 < x < 0.8$ already by earlier global analyses, the sea quark
and gluon polarization could only, still poorly, be constrained with the inclusion of  
information from SIDIS and $pp$ scattering data, as shown in Fig.~\ref{DSSV-full-set} 
(left panels).

The impact of SIDIS data is greatly illustrated with the extraction of the strange quark 
polarization. 
Figure~\ref{DSSV-full-set} (right) shows the $\chi^2$ profiles for the partial contributions 
($\Delta \chi^2$) of the various types of data sets, which clearly 
demonstrates that the information from SIDIS kaon production provides the
most stringent constraints on the shape of the strange helicity distribution.
The newly derived strange polarization is consistent with recent results from 
HERMES~\cite{Airapetian:2008qf} and COMPASS~\cite{Alekseev:2010ub} experiments, 
but exhibits a shape that is very different from those distributions obtained 
in earlier global NLO extractions, which 
employed constraints based on the assumption of SU(3) symmetry.
The polarization of strange quarks has been a focus since the very beginning of the 
'proton spin crises' and our poor knowledge so far
emphazises the need for much more precise data of semi-inclusive kaon production 
over a wide kinematic range.

\subsection{Fragmentation functions}

Very precise and clean information on parton-to-hadron fragmentation functions is
provided by data from electron-positron annihilation into charged hadrons.
Such data, however, do not allow to disentangle quark from antiquark fragmentation
without employing model dependent so-called 'tagging' techniques. 
Also gluon fragmentation is only poorly determined because of lack of precise enough 
data at energy scales away from the $Z$-resonance.
As for PDFs, complementary information on the flavour dependence of the fragmentation process 
can be obtained from single-hadron production in $pp$ collisions and semi-inclusive DIS.
These new data together with very precise PDF-sets allow for direct flavour tagging
of fragmentation functions.

\begin{figure}[t]
\begin{center}
\includegraphics[width=.8\textwidth]{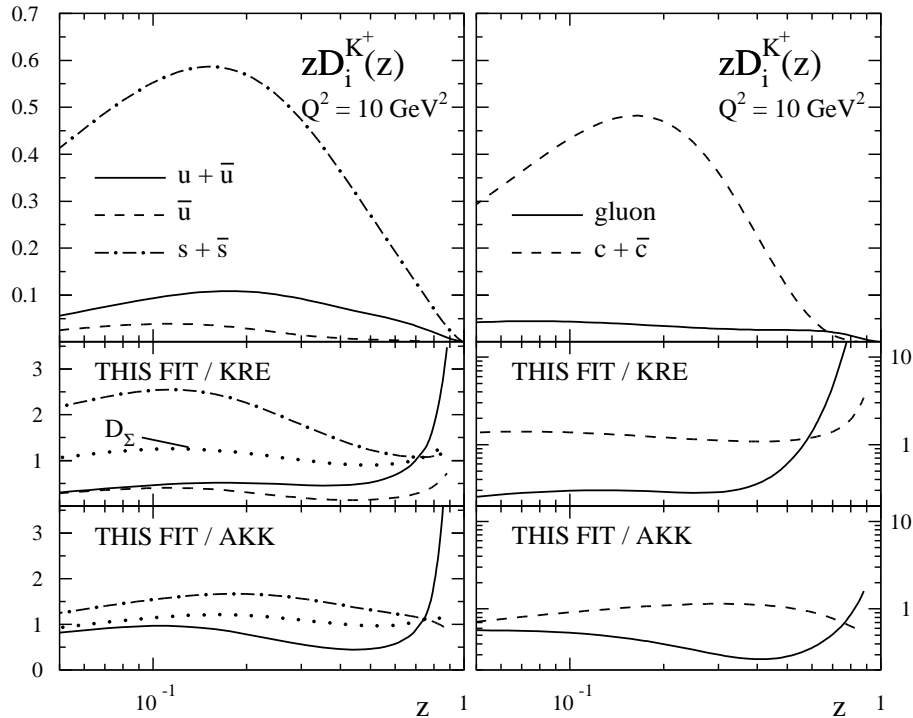}
\caption{\small 
Upper panels: the DSS individual fragmentation functions for positively
charged kaons $zD_i^{K^{+}}$
for $i=u+\bar{u},\, \bar{u},\, s+\bar{s},\, g,$ and $c + \bar{c}$.
Middle panels: ratios of DSS fragmentation functions to the ones
of KRE~\cite{Kretzer:2000yf}. The dotted line indicates the ratio for singlet
combination of fragmentation functions $zD_{\Sigma}^{K^{+}}$, where $\Sigma$ represents
the sum over all quark flavours.
Lower panels: ratios of DSS fragmentation functions to the ones
of AKK~\cite{Albino:2005me}; note that $D_{\bar{u}}^{K^{+}}$ is not
available in the AKK analysis.
}
\label{DSS-kaon}
\end{center}
\end{figure}

Such a comprehensive global QCD analysis was performed by the 
DSS-group~\cite{deFlorian:2007aj,deFlorian:2007nk}. 
For the first time, these new pion and kaon fragmentation functions reproduce well the wealth of 
precise electron-positron annihilation data {\it and} electron-proton and proton-proton 
observables.
Figure~\ref{DSS-kaon} shows the DSS individual quark fragmentation functions for $K^+$ and 
compares the new set with earlier extractions by KRE~\cite{Kretzer:2000yf} and AKK~\cite{Albino:2005me}
based only on $e^+e^-$ annihilation data and {\it assumptions} for a flavour separation.
The new fragmentation functions are very distinct from the earlier ones in emphazising the role 
of strangeness in kaon production (but also in pion production, see Ref.~\cite{deFlorian:2007nk}).
This result is driven by the charged kaon data measured in semi-inclusive DIS by the 
HERMES experiment and in $pp$ scattering by the BRAHMS experiment.
In particular the SIDIS data rule out the flavour separation {\it assumed} for kaons in  
KRE~\cite{Kretzer:2000yf}.

In future, very precise data will become available on the production of identified
hadrons measured in semi-inclusive DIS by COMPASS and CLAS12 experiments and in $pp$ scattering 
at RHIC and the LHC. 

These new data sets will yield very precise individual parton-to-hadron fragmentation functions
derived without assumptions on flavour separation as pioneered by the DSS group.
They are an indispensable ingredient for future global QCD analyses of standard PDFs
and TMD distributions, the latter are discussed in Section~\ref{sec:transverse}.

\clearpage

\section{\label{sec:transverse}The Transverse Structure of the Nucleon}

Modern polarized DIS experiments, capable to also detect and identify hadrons in 
the final state, are now revealing the intrinsic richness of the nucleon.
They aim for exploring two complementary aspects of nucleon structure:
the distribution of partons in the transverse plane in momentum space and 
coordinate space, encoded in the transverse momentum dependent and generalized 
parton distributions, respectively.
These new functions, for the first time, provide a framework for obtaining a three-dimensional
picture of the nucleon.
The mapping of TMDs and GPDs is a highly complex task that calls for a comprehensive
program, combining new dedicated experiments with intense theoretical studies and lattice
QCD simulations.
The study of these novel parton distributions is a major focus of the CLAS12 physics 
program. 
In this chapter we will discuss TMD distribution and fragmentation functions 
and highlight the particular role strangeness may play.

\subsection{Introduction}

Parton distribution functions, as introduced in the previous chapter, depend
also on the intrinsic parton momentum component $\boldmath{k}_\perp$ transverse to that 
of the nucleon: 
$f(x,k_\perp)$,
so-called 
transverse momentum dependent distributions (or short TMDs)~\cite{Mulders:1995dh,Bacchetta:2006tn}.
The study of TMDs in semi-inclusive DIS was put on a firm theoretical basis with the 
factorization proof for the kinematic regime where the transverse momentum of the produced hadron
is much smaller than the hard scale of the process
$P_{hT}^2 \ll Q^2$~\cite{Collins:1981uk,Ji:2004wu}.

Historically, 
transverse momentum dependent distribution or fragmentation functions have first been suggested
to explain the surprisingly large and otherwise puzzling 
single-spin asymmetries observed in hadronic reactions with transversely polarized protons.

More recently,
TMDs (as well as the GPDs discussed in the next chapter) have received much attention in the context
of the so-called 'spin puzzle':
the spin of quarks and gluons accounts only for a part of the nucleon spin. 
A substantial fraction of the nucleon spin must be due to orbital angular momentum. 
An intriguing aspect of certain TMDs is that nonzero values of these distributions require the 
presence of nucleon wave function components with different orbital angular momentum.
They hence provide information, in a model dependent way, on the elusive parton orbital motion 
and on spin-orbit effects.

%
\begin{table}[b]
\begin{center}
\begin{tabular}{|c|c|c|c|} \hline\hline
N $\backslash$ q & U & L & T \\ \hline
 {U} & ${\color{blue} \bf f_1}$   & & ${\color{red} h_{1}^\perp}$ \\
\hline
{L} & &${\color{green} \bf g_1}$ &    ${\color{red} h_{1L}^\perp}$ \\
\hline
 {T} & ${\color{blue} f_{1T}^\perp} $ &  ${\color{green} g_{1T}}$ &  ${\color{red} \bf h_1}$ \, ${\color{red} h_{1T}^\perp }$ \\
\hline\hline
\end{tabular}
\end{center}
\caption{\small 
Leading-twist transverse momentum dependent parton distributions. 
$U$, $L$, and $T$ stand for 
unpolarized, longitudinally polarized, and transversely polarized nucleons (rows) and 
quarks (columns).
}
\label{tab1} 
\end{table}
%
\vspace*{0.3cm}
There are eight leading-twist quark TMDs as summarized in Table~\ref{tab1}.
The three TMDs highlighted in boldface survive integration over $\boldmath{k}_\perp$.
These are the unpolarized (polarization averaged) momentum distribution $f_1(x,k_\perp)$ and the helicity 
distribution $g_1(x,k_\perp)$, discussed in the previous chapter, as well as the 
transversity distribution $h_1(x,k_\perp)$, which is related to transverse polarization of the 
struck quark.
The other five distributions in Table~\ref{tab1} do {\it not} survive integration over $\boldmath{k}_\perp$.
They typically describe the correlation between the transverse momentum of 
quarks, their spin and/or the spin of the nucleon, i.e. spin-orbit correlations.

Three of these TMDs, denoted by the letter $h$, describe the distribution of transversely polarized
partons.
In the helicity basis for the spin $\frac12$ nucleon, where $f_1$ and $g_1$ have their well known 
probabilistic interpretation, transverse polarization states are given by linear combinations of
positive and negative helicity states. 
Since helicity and chirality are the same at leading twist, the "$h$'' TMDs are therefore called 
{\it chiral-odd} distributions.
This peculiar property excludes them from influencing any inclusive DIS observable.
Chiral-odd TMDs appear in only those observables involving two chiral-odd partners, such as
Drell-Yan processes involving two chiral-odd parton distributions or semi-inclusive DIS 
involving also a chiral-odd fragmentation function.

Two TMDs, the Sivers $f_{1T}^\perp$~\cite{Sivers:1989cc} and the Boer-Mulders $h_{1}^\perp$~\cite{Boer:1997nt}
distributions,  
are rather exotic in being naive-time-reversal-odd (short: T-odd)\footnote{Naive time reversal involves the 
time reversal of three momenta and angular momenta without interchange of initial and final states.}.
For a long time T-odd effects were believed to vanish due to time reversal invariance~\cite{Collins:1992kk}.
Recently it was shown that initial and finale-state interactions can produce T-odd effects without
violating T-invariance~\cite{Brodsky:2002cx,Collins:2002kn,Belitsky:2002sm}.
These T-odd distributions play a crucial role in our understanding of nucleon structure.
Their observation, already confirmed for the Sivers distribution,
is a clear indication of parton orbital motion and the presence of non-trivial phases from
initial or final state interaction, that survive in the Bjorken limit.
The origin and expected process dependence of these functions, challenging the traditional 
concept of factorization and universality of PDFs, are related to fundamental QCD effects. 
In fact, the symmetry properties of QCD require the Sivers and Boer-Mulders
distributions to appear with opposite sign in the expressions for DIS and 
Drell-Yan cross sections~\cite{Collins:2002kn}.
The experimental verification of this peculiar breaking of universality for T-odd TMD distributions,
exhibits an important test for the description of single-spin asymmetries within the framework of QCD.
Its invalidation would have profound consequences for our understanding of high-energy reactions 
involving hadrons.
The chiral-even Sivers distribution describes the correlation of the parton intrinsic motion 
with the nucleon spin,
while the chiral-odd Boer-Mulders distribution relates this intrinsic parton motion with its
own spin in an {\it un}polarized nucleon.
Hence, the latter distribution has the striking peculiarity that it might give unexpected spin 
effects even in unpolarized processes.

Similar correlations arise in the hadronization process.  One particular case 
is the $T$-odd chiral-odd Collins fragmentation function $H_1^{\perp}(z,P_\perp)$ 
\cite{Collins:1992kk} representing a correlation between the transverse polarization  
of the fragmenting quark and the transverse momentum $P_\perp$ the produced hadron
acquires in the fragmentation process.
The Collins function 
is of high importance for spin physics because 
it acts as a ``quark polarimeter'', but it is also interesting on 
its own because it allows for an exploration of spin and orbital degrees of freedom 
of the QCD vacuum.

\vspace*{0.3cm}
Over the last decade, measurements of azimuthal moments of hadronic cross sections in 
hard processes have emerged  as a powerful tool for probing nucleon structure
through transverse single-spin asymmetries.
Many experiments worldwide are currently trying to pin down various 
TMD effects through semi-inclusive DIS (HERMES at 
DESY~\cite{Airapetian:1999tv,Airapetian:2001eg,Airapetian:2002mf,Airapetian:2004tw,Airapetian:2005jc,Airapetian:2006rx,Airapetian:2009ti,Airapetian:2010ds}, 
COMPASS at CERN~\cite{Alexakhin:2005iw,Alekseev:2008dn,Alekseev:2010rw,Bradamante:2011xu}, 
CLAS and Hall A at Jefferson Lab~\cite{Avakian:2003pk,Avakian:2005ps,Osipenko:2008rv,Avakian:2010ae,Aghasyan:2011ha}) 
polarized proton-proton collisions (BRAHMS, PHENIX and STAR at RHIC
~\cite{Lee:2009ck,Arsene:2008mi,Adler:2005in,Chiu:2007zy,Adare:2010bd,Dharmawardane:2010zz,Adams:2003fx,Abelev:2007ii,Abelev:2008qb,Eun:2010zz}), 
and electron-positron annihilation (Belle at KEK)~\cite{Abe:2005zx,Seidl:2008xc} and 
Babar at SLAC~\cite{Garzia:2011}).

The JLab 12 GeV upgrade will provide the unique combination of 
high beam energy, intensity (luminosity) and polarization, the usage of polarized targets
and advanced detection capabilities necessary to study the transverse momentum 
and spin correlations in polarized semi-inclusive DIS processes for a variety of
hadron species.
Full hadron identification over the whole kinematic range will be the key to
explore in detail the flavour dependence of TMDs.

\begin{figure}[t]
\begin{center}
\centerline{\epsfxsize=9.0cm\epsfbox{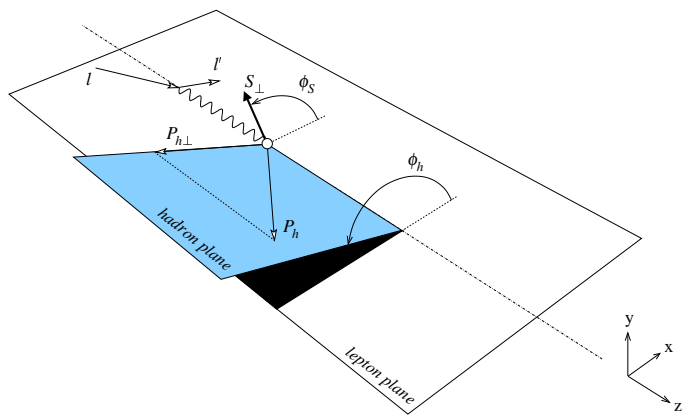}} 
\caption{\small 
Azimuthal angles in semi-inclusive DIS. For a longitudinally polarized target,
 $\phi_S$=0 or 180$^o$ for negative and
positive helicities of the proton, respectively.
}
\end{center}
\label{fig:anglestrento}
\end{figure}

\subsection{Spin and azimuthal asymmetries in semi-inclusive DIS}

When considering the nucleon internal transverse degrees of freedom, new mechanisms emerge that 
yield to azimuthal asymmetries in the distribution of observed hadrons.
Most generally, polarized single-hadron production in semi-inclusive DIS depends on six kinematic variables. 
In addition to the variables for inclusive DIS, $x$, 
$y$ (with $y = (P \cdot q) / (P \cdot l)$), and the azimuthal angle $\phi_S$ describing the
orientation of the target spin vector for transverse polarization,
one has three variables for the final state hadron, 
the longitudinal hadron momentum $z_h$, the magnitude of transverse hadron momentum $P_{hT}$, and
the azimuthal angle $\phi_h$ for the orientation of 
$P_{hT}$ (see also Fig.~\ref{fig:anglestrento}). 

The semi-inclusive DIS cross section can be written in a model independent way by 
means of a set of structure 
functions~\cite{Bacchetta:2006tn,Kotzinian:1994dv,Diehl:2005pc}.
At leading twist, this cross section is described by eight such structure functions, which
relate to different 
combinations of the polarization states of the incoming lepton 
and the target nucleon (for a complete expansion of the structure functions up to twist-3
see Ref.~\cite{Bacchetta:2006tn}).
Largely following the notation of~\cite{Bacchetta:2006tn}, one has
\begin{align}
\frac{d\sigma}{dx_B \, dy\, d\phi_S \,dz_h\, d\phi_h\, d P_{hT}^2}
\propto & \Big\{ F_{UU ,T} + \varepsilon \, \cos(2\phi_h) \,
F_{UU}^{\cos 2\phi_h} \nonumber \\ & + S_\parallel \, \varepsilon
\, \sin(2\phi_h) \, F_{UL}^{\sin 2\phi_h} + S_\parallel \,
\lambda_\ell \, \sqrt{1-\varepsilon^2}\, F_{LL} \phantom{\Big[}
\nonumber \\  & + | \mbox{\boldmath $S$}_\perp| \, \Big[
  \sin(\phi_h-\phi_S)\,
F_{UT ,T}^{\sin\left(\phi_h -\phi_S\right)} + \varepsilon \,
\sin(\phi_h+\phi_S) \, F_{UT}^{\sin\left(\phi_h +\phi_S\right)}
\nonumber \\ & \hspace{1.3cm} + \varepsilon \,
\sin(3\phi_h-\phi_S) \, F_{UT}^{\sin\left(3\phi_h -\phi_S\right)}
\Big] \nonumber \\  & + |\mbox{\boldmath $S$}_\perp| \, \lambda_e
\,
  \sqrt{1-\varepsilon^2} \, \cos(\phi_h-\phi_S) \,
F_{LT}^{\cos(\phi_h -\phi_S)} + \ldots \Big\} \,.
\label{e:crossmaster}
\end{align}
Here, $\varepsilon$ is the degree of longitudinal polarization of the
virtual photon which can be expressed through
$y$~\cite{Bacchetta:2006tn}, $\lambda_e$ is the lepton
helicity, and $S_\parallel$ denotes longitudinal and $S_\perp$
transverse target polarization.  The structure functions $F_{XY}$
merely depend on $x$, $z$, and $P_{hT}$.
Here, $XY$ refer to the lepton and the nucleon, respectively: $U$ =
unpolarized, $L, T$ = longitudinally, transversely polarized.
The third subscript $F_{XY,T}$ indicates the contribution from transversely 
polarized virtual photons.

By choosing specific polarization states and weighting with the
appropriate azimuthal dependence, one can extract each structure
function in~(\ref{e:crossmaster}).
These structure functions contain the information about all eight leading twist 
quark TMD distributions:
\begin{eqnarray} \label{eq:sf_tmd}
&&F_{UU,T} \sim \sum_q e_q^2 \> f_1^q \otimes D_1^q \hskip 54pt
F_{LT}^{\cos(\phi_h - \phi_S)} \sim \sum_q e_q^2 \> g_{1T}^{q}
\otimes D_1^q
\label{eq:sf_tmd1} \\
&&F_{LL} \, \sim \sum_q e_q^2 \> g_{1}^q \otimes D_1^q \hskip 60pt
F_{UT,T}^{\sin(\phi_h - \phi_S)} \sim \sum_q e_q^2 \> f_{1T}^{\perp q}
\otimes D_1^q
\label{eq:sf_tmd12} \\
&&F_{UU}^{\cos(2\phi_h)} \sim \sum_q e_q^2 \> h_{1}^{\perp q}
\otimes H_1^{\perp q} \hskip 24pt F_{UT}^{\sin(\phi_h + \phi_S)}
\sim \sum_q e_q^2 \> h_{1T}^{q} \otimes H_1^{\perp q}
\label{eq:sf_tmd3} \\
&&F_{UL}^{\sin(2\phi_h)} \sim \sum_q e_q^2 \> h_{1L}^{\perp q}
\otimes H_1^{\perp q} \hskip 24pt F_{UT}^{\sin(3\phi_h - \phi_S)}
\sim \sum_q e_q^2 \> h_{1T}^{\perp q} \otimes H_1^{\perp q} \,,
\label{eq:sf_tmd4}
\end{eqnarray}
where $e_q$ is the charge of the struck quark in units of the
elementary charge
and the symbol $\otimes$ denotes a convolution integral over intrinsic and 
fragmentation transverse momenta.
Here, all distribution and fragmentation functions depend also on the quark transverse 
momenta before and after scattering $k_{\perp}$ and $P_{\perp}$, respectively, 
$f(x,k_\perp)$ and $F(z,P_\perp)$, as well as on the scale $Q^2$.
To date, the transverse momentum dependence of TMDs is still largely unmeasured, and the 
common assumption of a Gaussian distribution is used in order to resolve the convolution
integral.
Very little is known about the average intrinsic transverse momentum that defines
the Gaussian width, or its flavour and kinematic dependence. 
In order to pin down such dependences, fully differential measurements of TMD related 
observables are highly demanded.

\begin{figure}[t]
\begin{minipage}[b]{6.2cm}
\epsfig{file=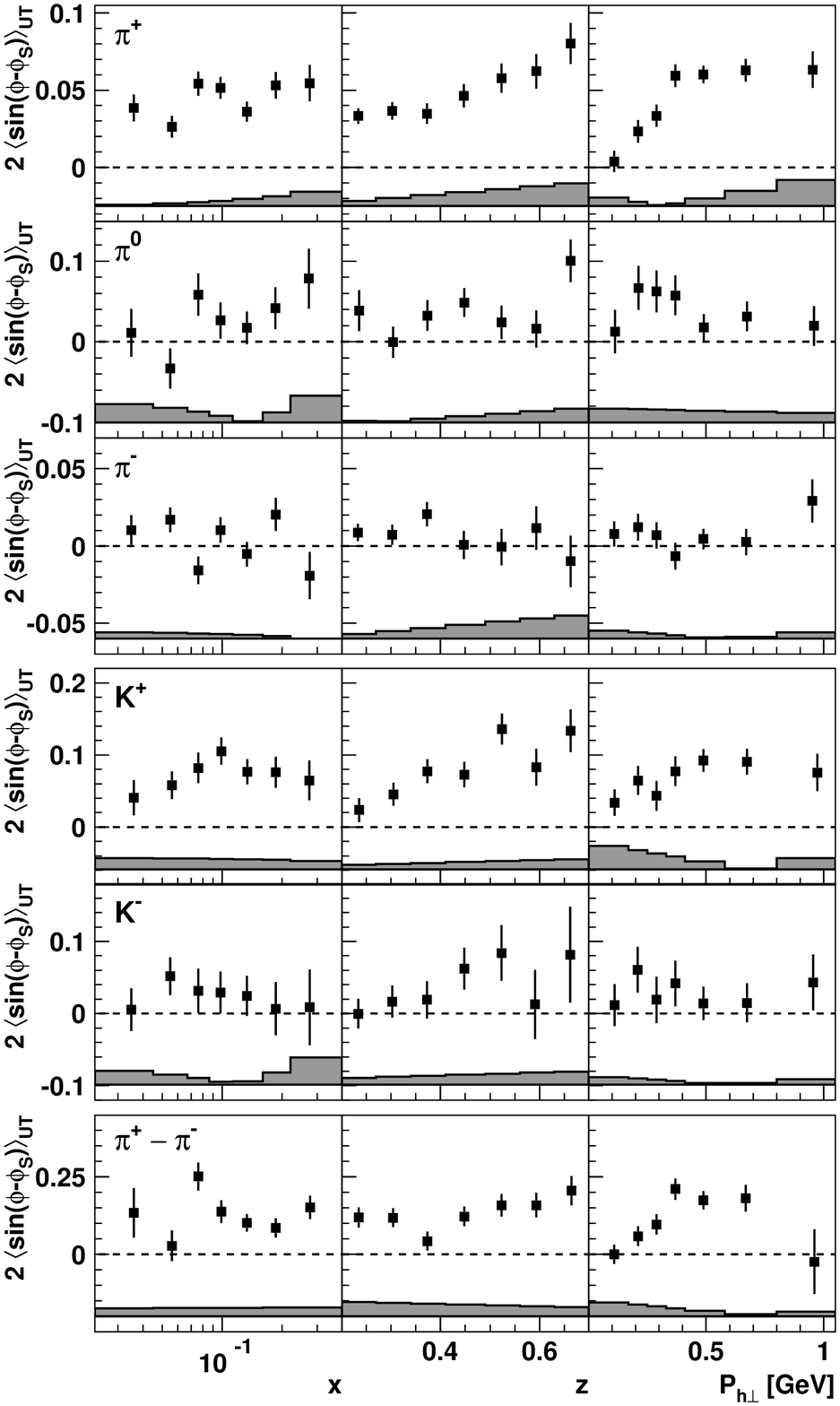,width=7.5cm,height=9.5cm}
\end{minipage}
     \ \hspace{5mm} \hspace{5mm} \
\begin{minipage}[b]{6.2cm}
\epsfig{file=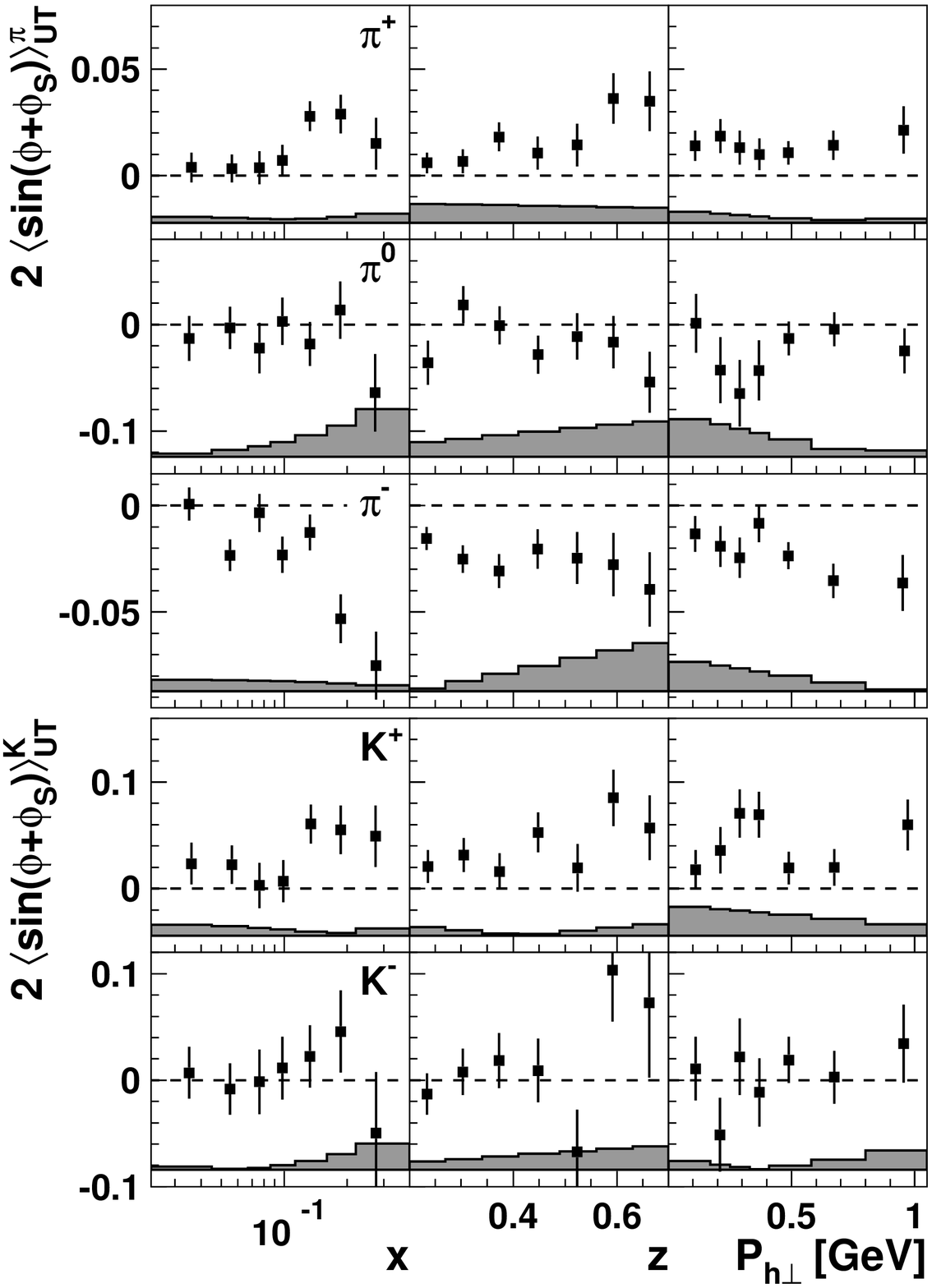,width=7.5cm,height=9.5cm}
\end{minipage}
\caption{ \small 
The Sivers (left) and Colllins (right) asymmetries for pions and kaons 
(as indicated in the panels) extracted
by the HERMES Collaboration~\cite{Airapetian:2009ti,Airapetian:2010ds}.}
\label{fig:hermeskaons}
\end{figure} 

Over the last few years, first results on transverse single-spin asymmetries 
became available from semi-inclusive DIS experiments.
Pioneering measurements by the HERMES Collaboration 
indicated significant azimuthal moments, which, for the first time, relate unambiguously 
to the Collins $(F_{UT}^{\sin(\phi_h+\phi_S)})$ and Sivers ($F_{UT}^{\sin(\phi_h-\phi_S)}$) 
effects~\cite{Airapetian:2009ti,Airapetian:2010ds}.
As discussed in the following section, striking differences were observed between charged pions and kaons 
for all TMD observables that are nonzero, indicating a surprisingly significant role sea quarks
may play in the transverse structure of the nucleon.
Thus measurements with identified strange hadrons over a wide kinematic range will provide the 
necessary information for clarifying the role of sea quarks and strangeness in particular.

\subsubsection{\label{sec:kaon-puzzle}The kaon ``puzzle''}

Figure~\ref{fig:hermeskaons} presents recent HERMES measurements for the Sivers (left) and
Colllins (right) asymmetries for pions and kaons~\cite{Airapetian:2009ti,Airapetian:2010ds}.
A very interesting facet of the data 
is the magnitude of the $K^+$ asymmetries, which is nearly twice as large as that
of $\pi^+$. 
As scattering of $u$ quarks dominates these data due to the charge factor, one might
naively expect that the $K^+$ and $\pi^+$ asymmetries should be similar.
Their difference in size may thus point to a significant role of other quark flavors, 
e.g. sea quarks.
This difference in magnitude for  $K^+$ and $\pi^+$ asymmetries is most pronounced for $x$ values
around 0.1 and larger, a kinematic region which will be explored in great detail
with CLAS12.

\begin{figure}[t]
\mbox{
\hspace*{-0.3cm}
\includegraphics[width=0.49\textwidth]{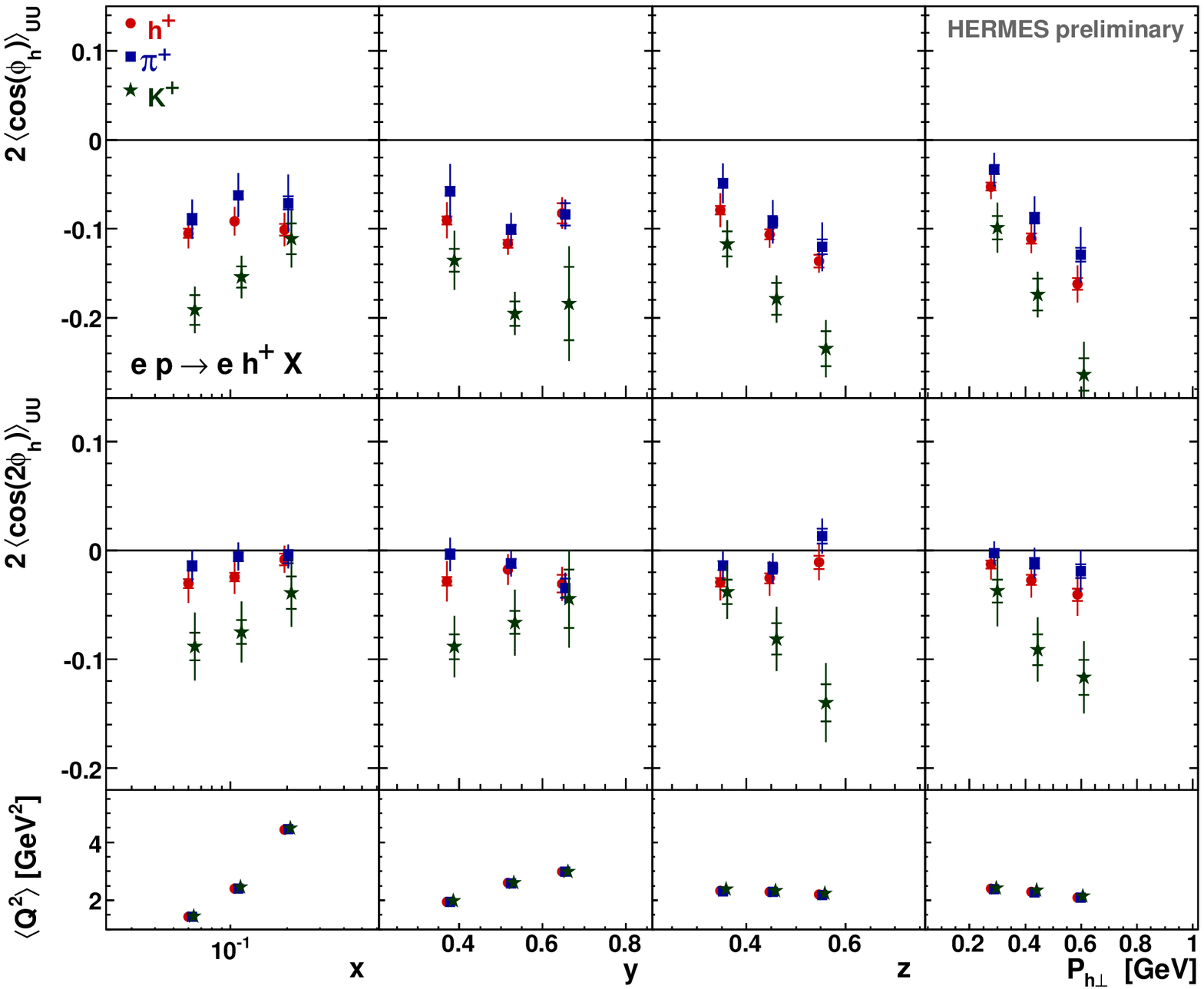}
\includegraphics[width=0.49\textwidth]{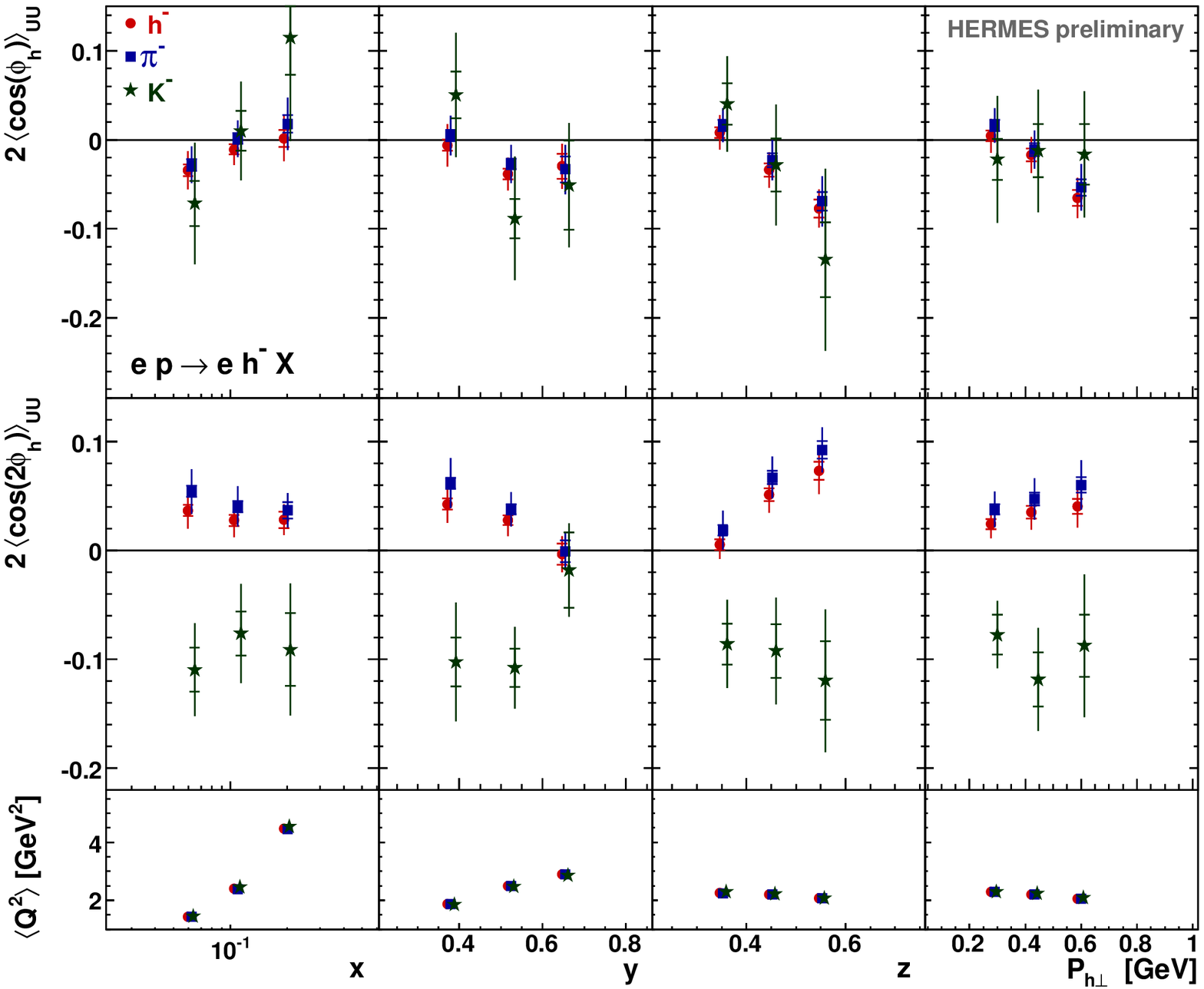}
}
\caption{\small 
The  $\cos\phi$ (upper row of each panel) and $\cos 2\phi$ (lower row of each panel) asymmetries 
for positively (left) and negatively (right) charged hadrons 
of different types as indicated in the panels, 
extracted by the HERMES Collaboration~\cite{Giordano:2010gq}.
}
\label{hermespi-K}
\end{figure}

Similarly, a very distinct pattern for pion and kaon asymmetries was also found 
when measuring the azimuthal 
dependence of the {\it un}polarized cross section, as shown in Fig.~\ref{hermespi-K}~\cite{Giordano:2010gq}.
For an unpolarized target the only azimuthal asymmetry
arising at leading twist is the $\cos2\phi_h$ modulation in Eq.~\ref{e:crossmaster}, which relates to
the Boer-Mulders distribution $h_{1}^\perp$ 
\begin{equation}
F^{\cos2\phi_h}_{UU} \propto \sum_{q} e^2_q ( h^{\perp q}_{1} \otimes H_1^{\perp q} + 
\frac{C}{Q^2} f_1^q  \otimes D_1^q + ... )
\label{eq:cos2phi}
\end{equation}
where $C$ is a kinematic factor.
The second term in the r.h.s. side of Eq.~\ref{eq:cos2phi} represents the Cahn effect~\cite{Cahn:1978se} 
which contributes at kinematic twist-4 level.
In fact, the $\cos\phi_h$ asymmetry shown in the upper row of the panels in Fig.~\ref{hermespi-K}
arises from the Cahn effect and the Boer-Mulders function which here both appear at sub-leading 
twist (twist-3). 
The Cahn effect accounts for the parton intrinsic
transverse momenta in the target nucleon ($\boldmath{k}_\perp$)  and the fact that produced hadrons 
may acquire transverse  momenta during the fragmentation process ($P_\perp$), so that the
final hadron transverse momentum $P_{hT}$ is defined as their sum $P_{hT} = z k_\perp+P_\perp \,$.
Theoretical estimates of this effect are still plagued by large uncertainties, mainly
related to the insufficient knowledge of the transverse momentum dependence of 
$f_1$ and $D_1$.
Moreover, the average intrinsic transverse momentum $k_\perp$ might depend on the quark flavour;
thus a flavour-dependent measure of the Cahn effect in semi-inclusive DIS of identified hadrons
is highly desirable.

Looking in more detail at the 
HERMES measurements presented in Fig.~\ref{hermespi-K},
we note, apart from the significantly bigger magnitudes of the $K^+$ asymmetries, 
the opposite sign of the $\cos 2\phi$ asymmetry for negative kaons compared to 
negative pions.  
This distinct pattern hints at a substantial contribution from sea quarks, in particular
from strange quarks.
This picture is in agreement with the large magnitude of $K^+$ amplitudes (larger than the $\pi^+$ ones)
for the Collins asymmetries $\langle \sin(\phi+\phi_S) \rangle _{UT}$ presented in Fig.~\ref{fig:hermeskaons}.
Both observations suggest consistently a Collins effect that is larger for kaons than for pions.
In particular, the difference with respect to pions can be related to a significant contribution
of {\it strange} quark to kaon production.

An extraction of the  Boer-Mulders distribution from semi-inclusive DIS is particularly exciting in view of existing 
measurements of this distribution from Drell-Yan 
processes~\cite{Falciano:1986wk,Guanziroli:1987rp,Zhu:2006gx}:
the observed violation of the Lam-Tung relation~\cite{Lam:1978zr} is a substantial hint for the
Boer-Mulders effect~\cite{Boer:1999si}.
Being a T-odd distribution, just like the Sivers function, the Boer-Mulders function also involves
a reversal in sign when going from DIS to Drell-Yan.
Experimental verification of this direct prediction of QCD remains outstanding.

\subsubsection{Hunting for the Collins function for kaons}

The chiral-odd Collins fragmentation function, discussed in the introduction, is an ideal and
necessary
partner for measuring chiral-odd distributions like transversity and Boer-Mulders distribution 
in semi-inclusive DIS.  
This function can be extracted from azimuthal asymmetries in the distribution of
back-to-back hadrons in two-jet events in electron-positron annihilation~\cite{Boer:1997mf}.
Pioneering measurements of these asymmetries for pions
have been performed by the BELLE Collaboration (KEK)~\cite{Abe:2005zx,Seidl:2008xc}.
First global analyses of semi-inclusive DIS and and $e^+e-$ annihilation data yielded recently the milestone 
extraction~\cite{Anselmino:2007fs,Anselmino:2008jk} of 
transversity for up and down quarks and the $u$ and $d$ to pion Collins fragmentation functions.
They revealed a very peculiar feature of the Collins FF: the {\it un}favoured function being of 
similar magnitude and opposite sign than the favoured
\footnote{Here, favoured (unfavoured) refers to the case where the fragmenting quark
appears (does not appear) as a valence quark in the produced hadron, e.g. the 
fragmentation of an up quark into a $\pi^+$ ($\pi^-$).} 
one.

However, no experimental information  is available yet about the
Collins fragmentation function for {\it kaons}. 
Model calculations of the Collins function for kaons~\cite{Bacchetta:2007wc}
 indicate that it might be comparable with pion Collins functions. 
Assuming isospin and charge-conjugation relations, 
there are in principle seven independent Collins fragmentation functions, 
but based on the observation that the pion favored Collins function is roughly
equal and opposite to the disfavored one, the number of independent Collins
functions could be reduced to three.
One could then access independent information about the Collins function from semi-inclusive DIS by
measuring asymmetries built from the differences between $\pi^+$ and $\pi^-$  and 
$K^+$ and $K^-$ using hydrogen and deuterium targets.
For example, assuming the strangeness distribution being negligible 
in the valence kinematic region at sufficiently high values of $x$, one can built the following ratio:

\be
 \frac{H_1^{u/K+}- H_1^{u/K-}}{H_1^{u/\pi +}- H_1^{u/\pi -}}  \propto
\frac{F_p^{K+} - F_p^{K-}}{3(F_p^{\pi +}-F_p^{\pi -}) + (F_d^{\pi +}-F_d^{\pi -})},
\ee

\noindent where $F_{target}^{hadron}$ can be any one of four asymmetries
related to $H_1^\perp$, like $\la \cos2\phi\ra_{UU}$,  $\la \sin2\phi \ra_{UL}$,  
$\la \sin(\phi+\phi_S) \ra_{UT}$ or $\la \sin(3\phi-\phi_S) \ra_{UT}$. 
Such ratios of asymmetries, simultaneously measured for pions and kaons,
will provide 
independent information about the Collins function for global analyses of TMDs.

\subsubsection{Dihadron production in semi-inclusive DIS and transversity}

The quark transversity distribution, $h_1$, as discussed in the introduction, is one of the
three basic quark distributions, together with $f_1$ and $g_1$, that survive $k_\perp$ integration.
Experimentally, transversity is quite an elusive object as its chiral-odd nature
excludes it from influencing any inclusive observable, which is our main source
of information about $f_1$ and $g_1$.
It needs to be measured together with another chiral-odd partner.
As discussed in previous sections, a very promising way is semi-inclusive DIS 
involving the chiral-odd transverse momentum dependent Collins fragmentation function in $F_{UT}^{\sin(\phi+\phi_S)}$, 
as exemplary presented in Fig.~\ref{fig:hermeskaons} (right). 

An interesting, alternative approach is semi-inclusive {\it dihadron} production from DIS, 
$ep^{\uparrow} \to e' (h_1\,h_2 ) X$, where the two unpolarized hadrons
emerge from the fragmentation of the struck quark.
The chiral-odd partner
of $h_1$ is represented by the chiral-odd Dihadron
Fragmentation Function (DiFF) $H_1^{\open q}$~\cite{Collins:1993kq,Bianconi:1999cd}.
Such measurements provide complementary information about transversity as here the
transverse spin of the fragmenting quark is transferred to the \emph{relative}
orbital angular momentum of the hadron pair. 
Consequently, this mechanism does not require transverse momentum of the hadron
pair and standard \emph{collinear} factorization applies.

There is much progress in the field of dihadron fragmentation from both experiment and theory.
Very recently, results from pioneering measurements of 
azimuthal correlations of two pion pairs in back-to-back jets 
in $e^+e^-$ annihilation related to the dihadron fragmentation function became
available from the BELLE Collaboration~\cite{Vossen:2011fk}.
On theoretical side, the dihadron cross section was computed
up to next-to-leading twist~\cite{Bacchetta:2003vn}  
and model calculations for the dihadron fragmentation functions are progressing.

The kinematics of the process is similar to the one in single-hadron semi-inclusive DIS
except for the final hadronic state, where now $z=z_{h1}+z_{h2}$ is the fractional energy carried by 
the hadron pair and the vectors $P_h=P_1+P_2$ and $R=(P_1-P_2)/2$ are introduced
together with the pair invariant mass $M_h$, which must be considered much smaller than the hard
scale (i.e., $P_h^2 = M_h^2 \ll Q^2$).

Similarly to Eq.~\ref{e:crossmaster}, the cross section for two-hadron production in 
semi-inclusive DIS can be expressed by structure functions.
When integrating over transverse momenta, only three structure functions appear at 
leading twist (the full next-to-leading twist expression is given in~\cite{Bacchetta:2003vn}) 
\begin{align}
\frac{d\sigma}{dx_B \, dy\, d\phi_R \,dz\, d\phi_S\, dM_h^2\, d\cos\theta}
\propto & \Big\{ F_{UU ,T} + S_\parallel \, \lambda_\ell \, \sqrt{1-\varepsilon^2}\, F_{LL} \phantom{\Big[}
\nonumber \\  
& + | \mbox{\boldmath $S$}_\perp| \, \varepsilon \,
\sin(\phi_R+\phi_S) \, F_{UT}^{\sin\left(\phi_R +\phi_S\right)} \Big\} \,,
\label{e:iff-cross-sec-leading-tw}
\end{align}
containing the three leading-twist distribution functions $f_1, g_1$ and $h_1$,
which now appear in a simple product with the involved fragmentation functions:
\begin{eqnarray}
&&\hspace*{-0.5cm}F_{UU,T} \sim \sum_q e_q^2 \> f_1^q(x_B) D_1^q(z,\cos\theta,M_h) \hskip 15pt
F_{LL} \sim \sum_q e_q^2 \> g_{1}^{q}(x_B)D_1^q(z,\cos\theta,M_h)
\label{eq:sf-dihad1} \\
&&\hspace*{-0.5cm}F_{UT}^{\sin\left(\phi_R +\phi_S\right)\sin\theta} \sim \sum_q e_q^2 \> h_{1}^{q}(x_B)
H_1^{\open q}(z,\cos\theta,M_h)  \,.
\label{eq:sf-dihad2}
\end{eqnarray}
The angle $\phi_R$ is the azimuthal angle of $R_T$, the component of $R$ transverse to 
$P_h$.  
A partial wave expansion reveals the structure of the dihadron fragmentation function.
For a two-hadron system with low invariant mass $M_h < 1.5$ GeV, 
dominant contributions come only from the lowest harmonics, i.e., the $s$ and $p$ waves.
The expansion in terms of Legendre functions of $\cos\theta$, 
with the polar angle $\theta$ evaluated in the center-of-momentum frame of the 
hadron pair, yields the factorized expressions
\begin{align}
D_1 & \simeq  D_{1,UU} + D_{1,UL}^{sp} \cos\theta + D_{1,LL}^{pp} \, \frac{1}{4} (3\cos^2\theta -1) \\
H_1^{\open } & \simeq  H_{1,UT}^{\open\, sp} + H_{1,LT}^{\open \, pp} \cos\theta \,.
\end{align}
The functions on the r.h.s. depend now only on $z$ and $M_h$.
When applying the symmetrization $f(\theta) + f(\pi - \theta)$, all $\cos \theta$ terms
will vanish even if the $\theta$ acceptance is not complete but still symmetric about $\theta = \pi/2$.
The dominant term is then 
$H_{1,UT}^{\open \, sp}$, the component of $H_1^{\open}$ that is sensitive to the 
interference between the fragmentation amplitudes into hadron pairs in relative 
$s$ and $p$ wave states.
It allows to study the transversity distribution without the complications of 
solving convolution integrals over transverse momentum or issues of factorization 
and evolution.
Pioneering measurements of two-pion production from polarized semi-inclusive DIS by 
HERMES~\cite{Airapetian:2008sk,Gliske-phd:2011} 
and COMPASS~\cite{Martin:2007au,Wollny:2009eq,Braun:2012} and from $e^+e^-$ annihilation by 
BELLE~\cite{Vossen:2011fk},
provided the necessary information for a first extraction of transversity in the framework
of collinear factorization~\cite{Courtoy:2011ni}.

For a real breakthrough of this promising approach to transversity, much
more data over a wide kinematic range are needed.
As for all semi-inclusive DIS observables, a safely bias-free extraction is only possible with 
an analysis that is performed fully differential in all relevant kinematic variables.
In the case of polarized dihadron production, the cross section depends on seven 
kinematic variables, which calls for even higher statistics than in the single hadron
production case. 
Hadron identification over the full accessible momentum range is an essential prerequisite
for a full exploration of the flavour dependence of the transversity distribution.
Particularly interesting are the $KK$ and $K\pi$ combinations, which exhibit sharp vector 
resonances, like the $\phi$, in the considered low invariant mass range~\cite{Gliske-phd:2011}.
These channels are ideal tools for studying the $sp$-wave interference via $H_{1,UT}^{\open \, sp}$, 
which is the leading part of $H_{1,UT}^{\open}$ giving rise to nonzero observables.

\subsection{Spin and azimuthal asymmetries in $pp$ collisions}

Historically, the surprisingly large left-right asymmetries measured in
hadronic reactions with transversely polarized protons initiated the idea about
a transverse momentum dependence of quark distributions in polarized protons.
The observation of such asymmetries was frequently quoted as a puzzle or
challenge for theory.
In fact, for a long time, transverse single-spin asymmetries were assumed to be negligible in hard
scattering processes~\cite{Kane:1978nd}.
The work of~\cite{Sivers:1989cc} introduced a transverse momentum dependent
quark distribution, now termed the Sivers function,
which provides a mechanism for the observed asymmetries that does not vanish at high energies.

The pioneering measurements at FermiLab~\cite{Adams:1991rx,Adams:1991ru,Adams:1991cs,Adams:1994yu,Bravar:1996ki}
 of these large (up to 0.3-0.4 in magnitude)
transverse-spin asymmetries in inclusive
forward production of pions in $pp$ collisions
$ p^\uparrow p \rightarrow \pi + X$,
have been greatly confirmed by experiments
at RHIC (BNL) at much higher center-of-mass energies of up to $\sqrt{s} = 200$ GeV
~\cite{Adams:2003fx}.
A rich variety of single-spin asymmetries for identified hadrons ($\pi^{\pm},\pi^{0},K^{\pm},p,\bar{p}$)
measured over a wide kinematic range is now available from the Brahms, Phenix and Star experiments
at RHIC
~\cite{Lee:2009ck,Arsene:2008mi,Adler:2005in,Chiu:2007zy,Adare:2010bd,Dharmawardane:2010zz,Abelev:2007ii,Abelev:2008qb,Eun:2010zz}.
The results exhibit a general pattern:
sizable asymmetries are measured at forward-rapidity and for positive Feynman $x_F > 0.3$
which increase in magnitude with increasing $x_F$ and $P_{hT}$.
In contrast,
for negative  $x_F$ and at mid-rapidity all asymmetries are found to be
consistent with zero.
Also for the observables in $pp$ reactions, a distinct behaviour for pions, kaons and protons was 
observed~\cite{Lee:2009ck}.

Several mechanisms have been suggested to explain these asymmetries.
At large values of \(P_{hT}\) collinear factorization involving twist-3 observables can be applied.
An alternative approach using TMDs has been used to describe existing data.
Employing recent parameterizations of the Sivers function based on fits to DIS data,
a fairly successful description of the observed asymmetries for pion production 
could be obtained~\cite{D'Alesio:2007jt}.

However, 
any kind of (standard) TMD factorization for the description of $pp$ reactions failed
so far due to the presence of both initial and final state interactions. 
Only on the basis of precise DIS data, involving a detailed study of the spin- and
transverse momentum dependence of hadron production in semi-inclusive DIS,
the rich field of results from hadronic collisions might be
interpreted within the QCD framework.

\clearpage
\section{\label{sec:exclusive} Nucleon Imaging: GPDs and Exclusive Processes}
\subsection{Introduction}

Complementary information towards a genuine multi-dimensional space and momentum resolution 
of the nucleon structure is offered by the so-called Generalized Parton Distributions 
(GPDs)~\cite{Mueller:1998fv,Radyushkin:1996nd,Ji:1996nm}.
These new functions, containing  the non-pertubative, long distance physics of factorized 
hard exclusive scattering processes, encompass the familiar PDFs and nucleon form factors 
as kinematic limits and moments, respectively.
GPDs offer opportunities to study a uniquely new aspect of the nucleon substructure: 
the localization of partons in the plane transverse to the motion of the nucleon. 
As such they provide a nucleon {\it tomography}. 
The ability to describe longitudinal momentum distributions at a fixed transverse localization 
is a prerequisite for studying the so-called Ji-relation~\cite{Ji:1996ek}, 
which links a certain combination of GPDs to the total angular momentum of a parton in the nucleon. 
From this quantity, the still hunted orbital angular momentum of partons in the nucleon could possibly 
be extracted, a question of crucial importance for an understanding of the nucleon structure.

Measurements of hard exclusive processes, where the nucleon stays intact and the final state is fully observed, 
are much more challenging than traditional inclusive and semi-inclusive scattering experiments. 
These exclusive processes require a difficult full reconstruction of final state particles and their cross-sections 
are usually small, demanding high luminosity machines. 
Pioneering measurements have been performed at DESY (HERMES, H1 and ZEUS) and JLab (HallA and CLAS).
A detailed study of GPDs is one of the central topics of ongoing and near-future experiments at COMPASS and JLab.
Most Importantly, the JLab 12 GeV upgrade
provides the unique combination of high energy, high beam intensity (high luminosity) and 
advanced detection capabilities necessary for studying low rate exclusive processes.

The exclusive production of real photons, Deeply Virtual Compton Scattering (DVCS), appears to be the theoretically 
cleanest process for studying GPDs. 
Similar to inclusive DIS, this process however, does not provide direct flavour dependent information.
A flavour {\it tagging} of GPDs could be gained from hard exclusive production of mesons where the
quark content of the meson provides flavour dependent information, similar to semi-inclusive DIS.
Moreover, vector meson production provides information about gluon GPDs even in the kinematic regime of JLab12, 
which predominantly probes the valence quark region:  
while the exclusive electroproduction of $\rho^0$ and $\omega$ mesons receives contributions from both quark and gluon 
exchange, the production of $\phi$ mesons is expected to be dominated by gluon exchange.
This unique access to gluon GPDs and hence to the gluon total angular momentum in exclusive $\phi$-meson production,
underlines the need for kaon identification over the whole kinematic acceptance of CLAS12 also for the
measurement of exclusive processes.
Kaon identification would significantly enhance also the capabilities for measurements of hard exclusive 
electroproduction of strangeness.

\subsection{Hard exclusive meson production}

Generalized parton distributions depend upon four kinematic variables: 
the Mandelstam variable  $t=(p-p')^2$, which is the squared momentum transfer to the target nucleon in the 
scattering process with $p$ $(p')$ representing the initial (final) four-momentum of the proton;
the average fraction  $x$ of the nucleon's longitudinal momentum carried by the active parton throughout
the scattering process; half the difference in the change of the fraction  of the nucleon's longitudinal momentum
carried by the active parton at the start and the end of the process, written as the skewness variable  $\xi$;
and the square $-Q^2 = q^2$ of the four-momentum of the virtual photon that mediates the lepton-nucleon
scattering process.
In the Bjorken limit of $Q^2 \rightarrow \infty$ with fixed $t$, $\xi$ is related to the Bjorken
variable as $\xi\simeq x_B/(2-x_B)$.

Observables measured in DVCS and hard exclusive meson production provide constraints on GPDs via 
Compton Form Factors (CFF). 
Each CFF is a convolution of a hard scattering kernel with a GPD that describes a soft part of the 
scattering process.
For meson production, the meson distribution amplitude enters as a second soft part 
the factorized cross section.  
In this case, factorization was proven for longitudinal virtual photons only and an experimental 
separation of the $\sigma_L$ and $\sigma_T$ cross sections is desirable.
At high  $Q^2$, however, the longitudinal part of the cross section is expected to dominate as in this limit
$R=\sigma_T/\sigma_L\sim 1/Q^2$. 
Recent model calculations also take the transverse part of the cross section into 
account~\cite{Goloskokov:2005sd,Goloskokov:2007nt,Goloskokov:2008ib}
and successfully describe a series of different observables in exclusive meson 
production.
A more recent work emphasizes the very interesting possibility for an access to GPDs for strange quarks from 
exclusive kaon-hyperon production and highlights the role of transversity GPDs in pseudoscalar
meson production~\cite{Goloskokov:2011rd}. 
Large, up to 0.4 in magnitude, transverse single-spin asymmetries are predicted for 
the $K^+\Lambda$ and $K^+\Sigma^0$ channels for the typical kinematics of JLab12.
Also for measurements with longitudinally polarized targets or beam,
sizeable asymmetries of about 0.05 to 0.1 in magnitude are predicted from the same model
calculations~\cite{Goloskokov:2011rd}.
Much progress from theoretical side on GPD model calculations is expected for the near future.

\subsubsection{Exclusive $\phi$-meson production}

Exclusive electroproduction of $\phi$ mesons provides unique access to the gluon GPD and hence, 
involving the Ji-relation, to the total angular momentum of gluons.
Since the  $\phi$ meson contains mainly strange quarks with a very small admixture of quarks with other 
flavours, quark-exchange with the nucleon is expected to be suppressed and two- (or more) gluon
exchange to dominate. 
\begin{figure}[t]
\begin{center}
\includegraphics[width=0.49\textwidth]{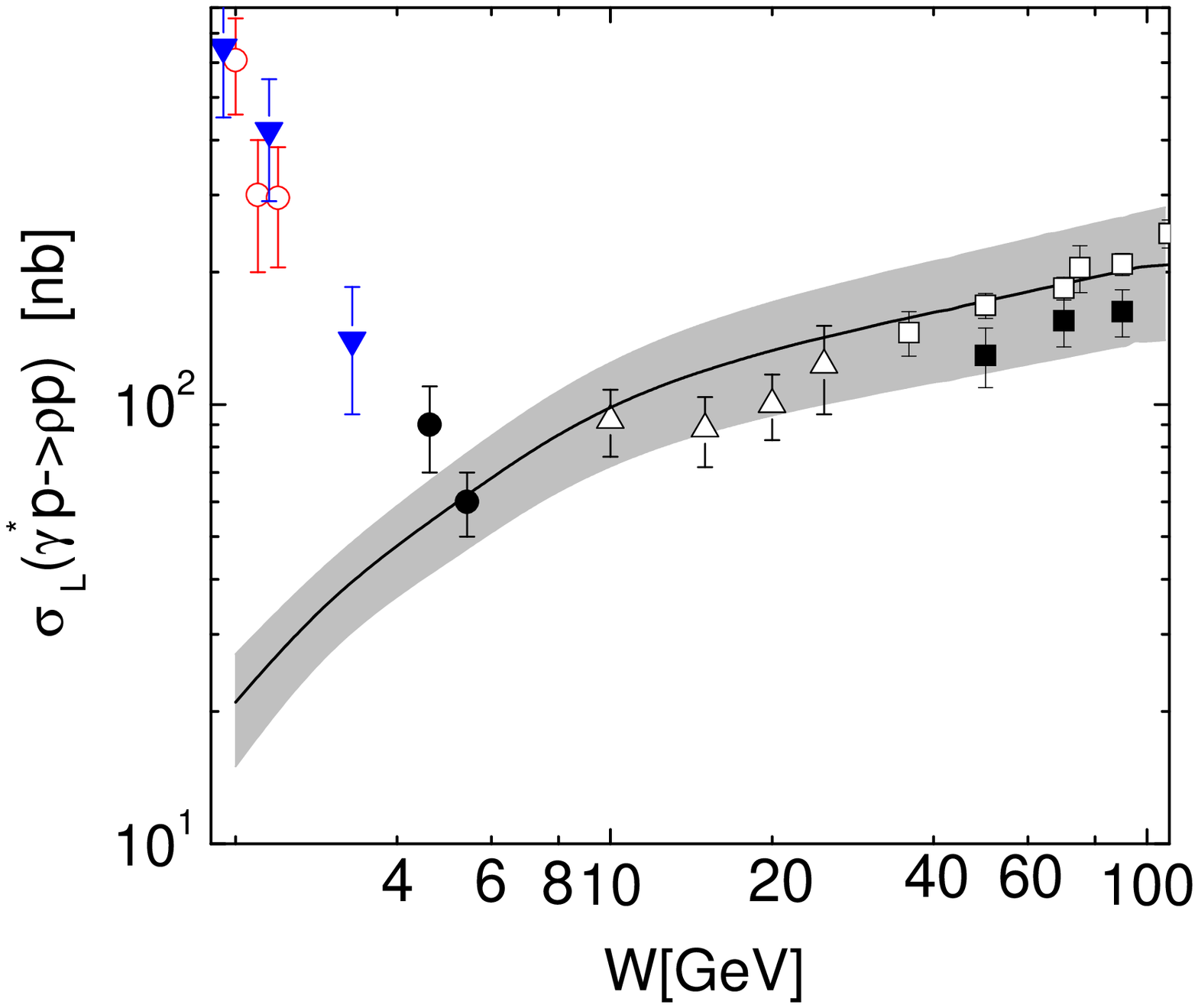}
\includegraphics[width=0.49\textwidth]{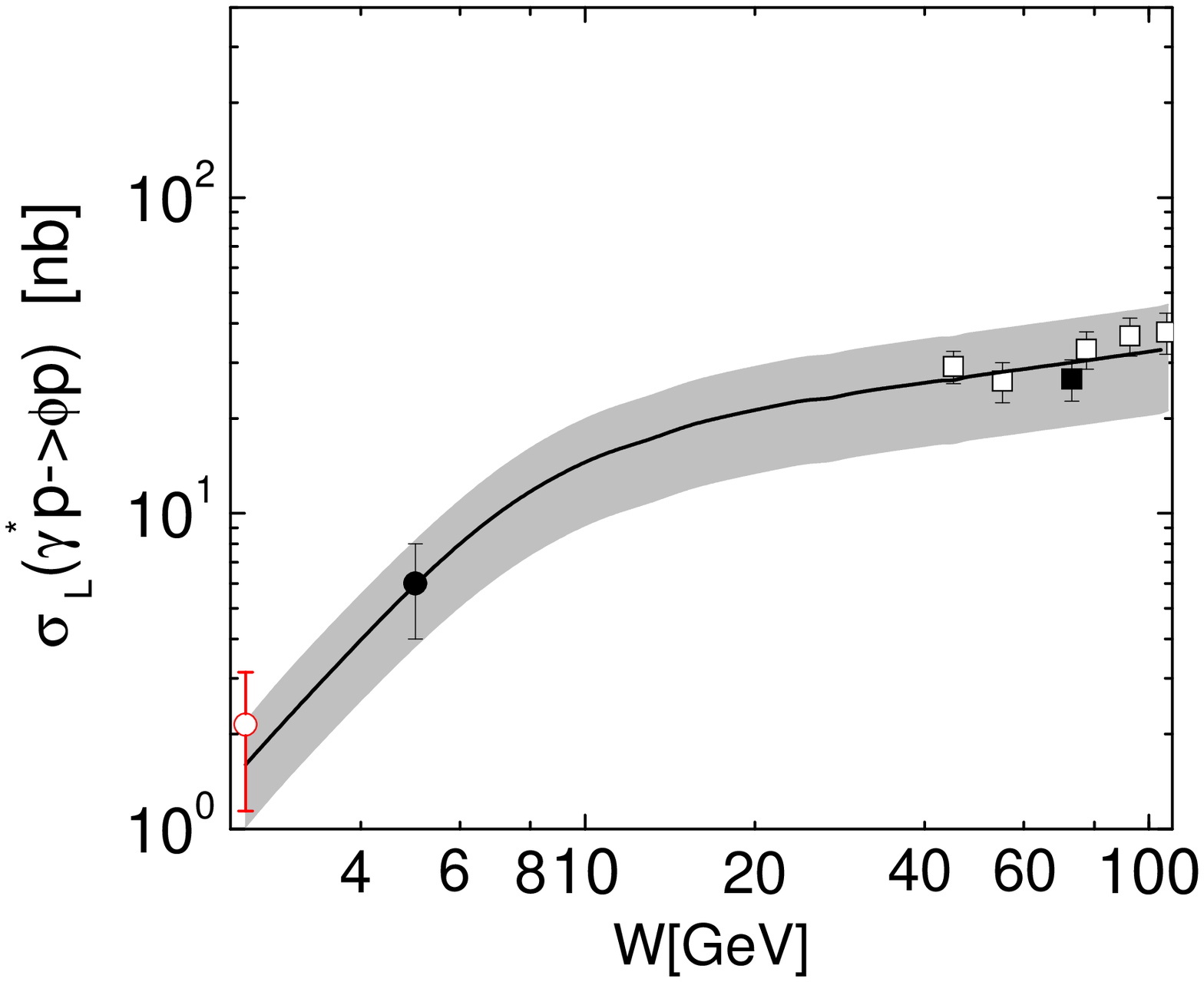}
\caption{\small Longitudinal part of the cross section as a function of $W$, for the exclusive production  of
$\rho^0$ (left) and  $\phi$ (right) mesons. 
The curve is  the result of a model calculation from~\cite{Goloskokov:2005sd,Goloskokov:2007nt,Goloskokov:2008ib}
with references for the presented data therein.
}
\label{phiGPDs}
\end{center}
\end{figure}
As shown in Fig.~\ref{phiGPDs},
available cross section data have been successfully described over a wide range in $W$, where $W$ is the 
invariant mass of the photon-nucleon system, using the GPD model of 
Ref.s~\cite{Goloskokov:2005sd,Goloskokov:2007nt,Goloskokov:2008ib}.
For the production of $\rho^0$ mesons, shown in the left panel of  Fig.~\ref{phiGPDs}, 
there is a clear indication for a change in the production mechanism around $W = 4$ GeV and the GPD 
model, where both quark and gluon exchange is taken into account, fails to describe the JLab data at low $W$.
In contrast, $\phi$-meson production is well described over two orders of magnitude in $W$ down to values 
of about 2 GeV by the employed GPD model, which assumes dominance of gluon exchange for this channel.
This underlines the usage of exclusive $\phi$-meson production as a very suitable channel for 
studying gluon GPDs and clearly much more data is needed over a wide kinematic range.

%
%
Experimentally, $\phi$-meson production is a very clean, nearly background free exclusive channel given the kaons 
from the decay $\phi\to K^+K^-$ can be identified.
A Monte Carlo simulation of the reaction  
$ep\to ep\phi$, $\phi\to K^+K^-$ obtained with 11 GeV electrons shows 
that the kaons from the $\phi$ decay are reaching momenta up to 6 GeV.
Hence, kaon identification over the momentum range 1--6 GeV would open the way for detailed measurements 
of this unique channel for studying the role of gluon GPDs in the valence kinematic region.

\clearpage

\section{\label{sec:nuclear}Hadronization in the Nuclear Medium}
\subsection{Introduction}

An interesting pattern of modifications of parton distribution and fragmentation functions
in nuclei has been observed, which caused much
excitement and vast experimental and theoretical activities.
Generally, the evolution of a fast-moving quark into hadrons is a non-perturbative, 
dynamic phenomenon and its space-time evolution is a basic issue of physics.
The understanding of quark propagation in the nuclear medium is crucial for the interpretation
of high energy proton-nucleus interactions and ultra relativistic heavy ion collisions, the
latter aiming for studying the Quark-Gluon Plasma.
Quark propagation in the nuclear environment involves different processes like 
multiple interaction with the medium and induced gluon radiation, which 
lead to attenuation and transverse momentum broadening of hadron yields compared to nucleon targets.

\subsection{SIDIS and hadronization in cold nuclear matter}

The hadronization process in ``free space'' has been studied extensively in electron-positron 
annihilation experiments. 
As a result, the spectra of particles produced and their kinematic dependences are rather well 
known (within the limitations discussed in Chapter~\ref{sec:longitudinal}).
In contrast, very little is known about the space-time evolution of the process.

Leptoproduction of hadrons has the virtue that the energy
and momentum transferred to the hit parton are well determined, as it is 
``tagged'' by the scattered lepton and the nucleus is basically used as 
a probe at the fermi scale with increasing size or density, thus acting as 
femtometer-scale detector of the hadronization process.
Theoretical models can therefore be calibrated in nuclear semi-inclusive DIS  and  
then applied, for example, to studies of the Quark-Gluon Plasma.

Effects of cold nuclear matter on hadron production in semi-inclusive DIS
have been extensively studied by the HERMES experiment
at DESY using a 27.6 GeV positron beam on internal
gaseous targets of deuterium, helium, neon, krypton or xenon~\cite{Airapetian:2011jp}. 
By using efficient particle identification over the whole kinematic range of the
interactions, the measurements have been performed for charged
and neutral pions, for charged kaons and for protons and anti-protons, 
exhibiting very distinct patterns for the different particle types.

The obtained results clearly reveal that the
capability of identifying pions, kaons and protons over the whole kinematic range
of interest is essential for gaining more insights into the details of the hadronization process.
Furthermore, the potential of performing a fully differential analysis is a key to disentangle the
various different stages of hadronization.

\subsubsection{Nuclear attenuation}

\begin{figure}[t]
\begin{center}
\mbox{
\includegraphics[width=0.48\textwidth]{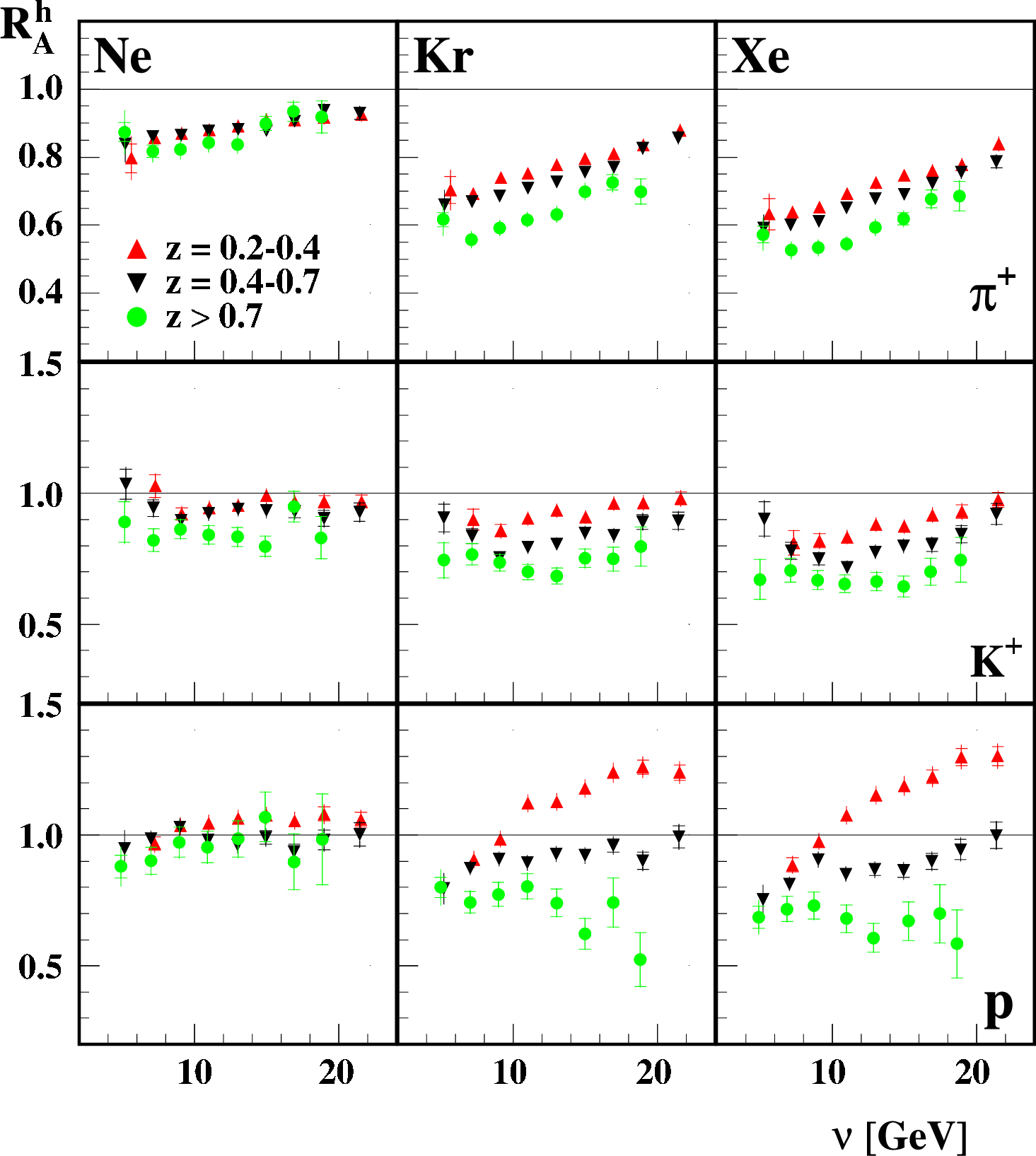}
\includegraphics[width=0.48\textwidth]{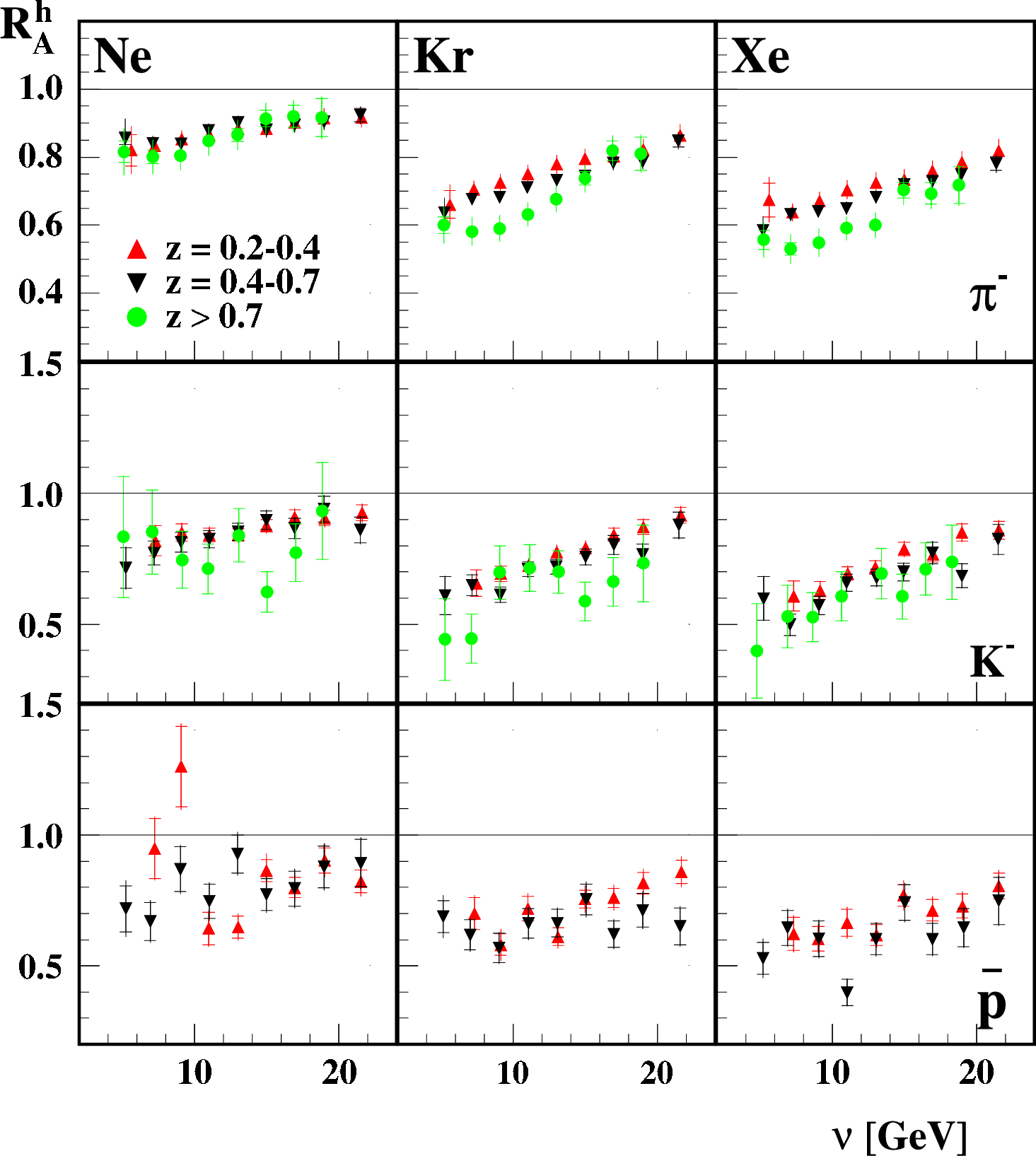}
}
\end{center}
\caption{\label{fig:atte} \small
The $\nu$ dependence of
the multiplicity ratio $R^h_A$ for positively (left)
and negatively (right) charged hadrons for three slices in $z$ as indicated in the legend,
measured by the HERMES Collaboration~\cite{Airapetian:2011jp}.
}
\end{figure}

The experimental results are usually presented in terms of the hadron multiplicity ratio
$R_A^h$, which represents the ratio of the number of hadrons of type $h$ produced per DIS event
on a nuclear target of mass $A$ to that from a deuterium target.
Here we only show a few examples that highlight the observed effects.
Figure~\ref{fig:atte} presents the multiplicity ratio $R^h_A$ as
a function of the virtual photon energy ($\nu$) for three slices
of the fractional virtual photon energy carried by the produced hadron ($z$).
A very striking observation is the behaviour of $R^h_A$ as a function
of  $\nu$ which indicates a clear attenuation at low $\nu$ values  
where the ratio is smaller than one and constantly increases
indicating smaller nuclear effects for high energy
transferred by the virtual photon. 
While this behaviour is similar for pions and kaons, it is 
very different for protons where knock-out
processes from the target remnant might contribute to the proton yield.
But even for pions and kaons, the magnitude of attenuation differs for the two
hadron types and the different charges.
While $\pi^+$ and  $\pi^-$ behave very similar, there are clearly visible differences
for  $K^+$ and $K^-$.
The various details of attenuation effects become especially pronounced when viewing their 
multi-dimensional kinematic dependences.
Deeper inside into the various different meachnisms causing nuclear effects,
would be gained with 
a fully differential analysis of this observable for identified pions, kaons and protons.

\subsubsection{Medium modification of the transverse momentum}

\begin{figure}[t]
\begin{center}
\includegraphics[width=0.5\textwidth]{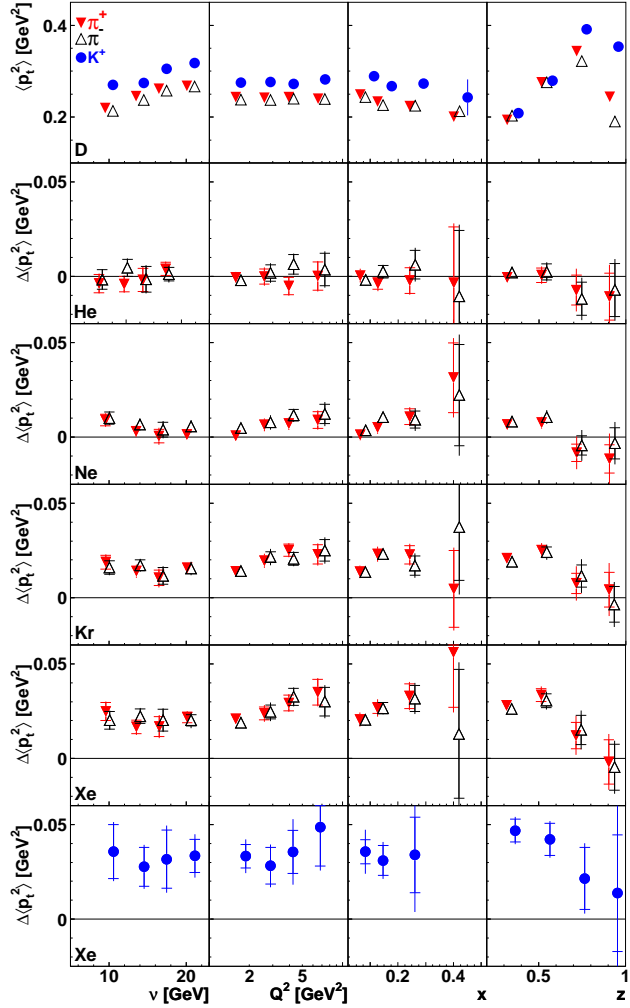}
\end{center}
\caption{\label{fig:pt} \small 
The average transverse momentum  $\langle p_t^2 \rangle$ for D (top row)
and $p_t$-broadening $\Delta \langle p_t^2 \rangle$ (remaining rows)
for $\pi^+$ and $\pi^-$ produced on He, Ne, Kr, and Xe targets and for $K^+$
produced on a Xe target (bottom row);
measured by the HERMES Collaboration~\cite{Airapetian:2009jy}.
}
\end{figure}
A complementary measurement to the hadron attenuation ratio, which is even
more sensible to the partonic stage of the hadronization process, is
the hadron transverse momentum broadening 
$\Delta \langle p_t^2 \rangle=\langle p_t^2 \rangle_A^h-\langle p_t^2 \rangle_D^h$. The first direct 
measurement has been performed by HERMES~\cite{Airapetian:2009jy}
and is shown in Fig.~\ref{fig:pt}.
The panels show the average transverse momentum $\langle p_t^2 \rangle$ for deuterium (top row) 
and the broadening $\Delta \langle p_t^2 \rangle$ (remaining
rows) 
as a function of either $\nu$, $Q^2$, $x$ and $z$ for $\pi^+$ or
$\pi^-$ for the various nuclear targets. The data do not reveal a significant
dependence on $\nu$ while for $x$ and $Q^2$ the effect slightly increases.
In contrast, the effect as function of $z$ shows a clear broadening, which is vanishing
as $z$ approaches unity. This indicates both that there is no or little 
dependence of the primordial transverse momentum on the size of the nucleus
and that the $p_t$-broadening is not due to elastic scattering
of pre-hadron or hadrons already produced within the nuclear volume.
There is a slight indication for larger  $p_t$-broadening effects for $K^+$ 
(bottom panels of Fig.~\ref{fig:pt}) compared to 
pions but much more statistics is needed to explore a possible hadron type dependence.

The interpretation of this observable would greatly benefit from both a 
fully differential analysis and full hadron identification.
For instance, a multi-dimensional study of the transverse momentum broadening in nuclear
matter, would provide information to distinguish the effects due to the primordial transverse momentum, 
gluon radiation of the struck quark, the formation and soft multiple interactions of 
the so-called ``pre-hadron'' and the interaction of the formed hadrons with the surrounding hadronic medium.

\subsubsection{Medium modification of TMD distributions}

\begin{figure}[t]
\begin{center}
\includegraphics[width=0.85\textwidth]{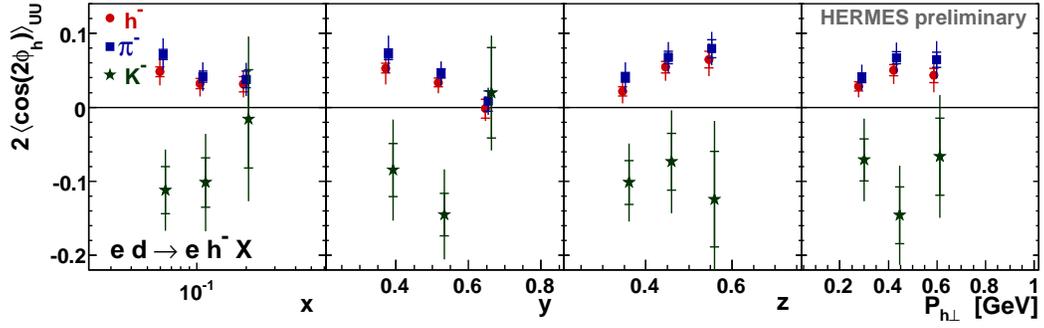}
\end{center}
\caption{\label{fig:cosphi-hermes}
\small
The $\langle \cos2\phi \rangle$ moments for negatively charged hadrons of types indicated in the
figure; extracted by the HERMES Collaboration~\cite{Giordano:2010gq} using a deuterium target.
}
\end{figure}

The study of nuclear effects also opens an alternative way to gain information about
certain TMD distributions discussed in Section~\ref{sec:transverse}. 
Azimuthal asymmetries like the $\langle \cos \phi \rangle$ and $\langle \cos 2\phi \rangle$ 
moments, are expected to be suppressed 
when being measured in nuclei due to final-state multiple scattering.
Model calculations~\cite{Gao:2010mj} demonstrated
a non-trivial transverse momentum dependence of this suppression  on the relative shape
of the involved TMD quark distributions.
Therefore, the nuclear modification of azimuthal asymmetries and the study of its transverse
momentum dependence is a very sensitive probe of TMD quark distributions.
As shown in  Fig.~\ref{fig:cosphi-hermes} for measurements with a deuterium target, the 
shape of the asymmetries for pions and kaons is quite different and hence, nuclear effects are
expected to be quite different for the various hadron types.
In order to observe such effects, however, measurements with nuclear targets much heavier than deuterium 
are needed.

\clearpage
\section{Spectroscopy: Hidden Strangeness and Strangeonia}
      The phenomenology of hadrons and in particular the study
      of their spectrum led more than forty years ago to the
      development of the quark model, where baryons and mesons
      are described as bound systems of three quarks and of a
      quark-antiquark pair, respectively. While this picture
      still holds and has been proven to reproduce many
      features of the hadron spectrum, now we know that the
      hadron mass cannot be explained only in terms of the
      quark masses, but it is mainly due to the dynamics of the
      gluons that bind them. Measuring the spectrum of hadrons,
      studying their properties and inner structure is
      therefore crucial to achieve a deep knowledge of the
      strong force.

      Mesons, being made by a quark and an anti-quark, are the
      simplest quark bound system and therefore the ideal
      benchmark to study the interaction between quarks,
      understand what the role of gluons is and investigate the
      origin of confinement. The quark model predicts the
      existence of multiplets of mesons with similar
      properties, that are classified according to their total
      angular momentum $J$, the parity $P$, and charge
      conjugation $C$. While most of the lowest mass states
      have been clearly identified and studied~\cite{Nakamura:2010zzi},
      several open issues related to the mass hierarchy and
      decays of excited states remain and still await for a
      thorough experimental investigation. In addition,
      phenomenological 
models~\cite{Jaffe:1975fd,Horn:1977rq,Barnes:1982tx,Isgur:1985vy,Isgur:1984bm,Guo:2008yz}
      and lattice QCD
      calculations~\cite{Dudek:2009qf,Dudek:2010wm} suggest
      that states beyond the simple $q\bar q$ configuration, such as
      hybrids ($qqg$), tetraquarks ($qq\bar q\bar q$) and
      glueballs, should also exist. If so, we should expect to
      find a much richer spectrum than that predicted by the
      quark model and, in particular, we should be able to
      observe new meson multiplets corresponding to these
      unconventional configurations. Hybrid mesons are of
      particular interest as they are the cleanest experimental
      signature for the presence of gluons in the dynamical
      mass generation process. A precise determination of their
      spectrum and properties can provide a unique opportunity
      to study the role of the glue and to understand the
      phenomenon of confinement.
      An unambiguous identification of these states is in
      general rather difficult, since they can mix with
      ordinary mesons having the same quantum numbers
      ($J^{PC}$). However, the additional degrees of freedom
      present in these states can also lead to {\it exotic}
      quantum numbers that are not allowed in $q\bar q$ systems
      and therefore provide a unique signature of their unusual
      structure. 
 
MesonEx is an experimental
      program to study meson spectroscopy via quasi-real
      photoproduction in Hall B, using the CLAS12 detector and
      a new Forward Tagger (FT) facility.
      The 12 GeV electron beam available after the upgrade of
      Jefferson Laboratory, the excellent characteristics of the
      CLAS12 detector and the new FT facility will give the
      possibility of exploring a broad mass range accumulating
      data of unprecedented accuracy and
      statistics.
      As shown in the Sections below, the MesonEx program will
      greatly benefit from a new RICH detector that will extend
      the kaon
      identification capabilities to large momenta of up to 7 GeV.

\subsection{Hybrids with hidden strangeness}
      One very attractive method to identify exotic mesons is
      through strangeness-rich final states, like the $\phi\pi$
      decay mode. Any $s\bar{s}$-meson decay to $\phi\pi$ is
      forbidden due to the conservation of isotopic spin.
      This
      decay mode is also forbidden for any light $q\bar{q}$-meson (with $q$ beeing a $u$ or $d$ quark) 
by the Okubo-Zweig-Iizuka (OZI) rule.
On the other hand, multiquark
      or hybrid mesons are expected to have a strong coupling
      to the $\phi\pi$ system. The discovery of a $\phi\pi$
      resonance would indicate a new kind of hadron and suggest
      a $q\bar{q}g$ or $q\bar{q}q\bar{q}$ state.
      This is true for $f'\pi$ and $J/\psi\pi$ decay modes as
      well~\cite{Close:1978be}.
      Some experimental evidence for the existence of a
      resonance with strong $\phi\pi$ coupling is available. In
      experiments at the LEPTON-F
      spectrometer~\cite{Bityukov:2005ag,Bityukov:1986yd}, a new meson
      C(1480), with mass $1480\pm40$~MeV, width $130\pm60$~MeV
      and an anomalous large branching ratio to $\phi\pi$, was
      observed.
      The angular distributions of the sequential decay
      $C(1480)\rightarrow\phi\pi^{0},\phi\rightarrow
      K^{+}K^{-}$ were studied and the quantum numbers for
      C(1480) meson have been determined to be
      $I^{G}=1^{+},J^{PC}=1^{--}$. 
For this meson, an anomalously
      large value of the ratio
$      BR(C(1480)\rightarrow\phi\pi^{0})/BR(C(1480)\rightarrow\omega\pi^{0})>0.5 $
      at 95\% C.L. was obtained. This value is more than two
      orders of magnitude higher than the expected ratio for
      mesons with the standard isovector quark structure.
      At present, the only consistent explanation of
      these properties can be obtained with the assumption that
      the $C(1480)$ meson is a four quark or hybrid state.
      At the $\Omega$--spectrometer~\cite{Atkinson:1983df} the cross
      section for the reaction $\gamma
      p\rightarrow\phi\pi^{0}p$ was measured. Although the
      number of events was not large ($\sim25$), an excess of
      events in the mass spectrum of the $\phi\pi^{0}$ system
      at $\sim$1.4 GeV was observed. The $\phi\pi^{0}$
      photoproduction cross section was estimated to be
      $\sigma(\gamma p\rightarrow\phi\pi^{0}p)=6\pm3{\rm nb}$
      (at 95\% C.L.).
      The existence of a structure in the same mass range was
      confirmed with the study of inclusive $\phi\pi^{+}$
      production with a pion beam~\cite{Antipov:1995kb} and by recent
      $e^{+}e^{-}\to\phi\pi^{0}$ data from Ref.~\cite{Aubert:2008jd}.

      A clean identification of the final state can be
      obtained by detecting the $K^+K^-$ pair from the charged
      decay of the $\phi$ meson. 
This could be achieved with the
      CLAS12 base equipment, where a RICH detector that
      extends kaon identification to momenta beyond 2.5 GeV, would
      greatly improve the accessible kinematic range for the higher 
momentum kaons produced in the reaction.
Such studies will open a unique
      window in the search of exotics with hidden
      strangeness.

\subsection{Strangeonia}

      Strangeonia are mesons containing $s\bar{s}$ pairs:
      these can be conventional states in the quark model or
      hybrids with or without exotic quantum numbers. While the
      strange meson spectrum is quite well understood, the
      details of the strangeonium spectrum are much less
      certain and only a handful of states have been confirmed.
      This, in itself, is a motivation for studying
      strangeonium spectroscopy. In addition, a number of the
      final states in which an exotic signal has been claimed
      are also final states in which strangeonia would be
      expected. Thus, a precise determination of the
      strangeonium spectrum is important to constrain the
      search of exotic candidates.

      The masses are expected to be in the 1-3 GeV range, i.e.
      a transition region between light (relativistic) and
      heavy (non-relativistic) $q\bar{q}$ states\footnote{The
      relevance of this aspect was pointed out by Barnes, Page
      and Black~\cite{Barnes:2002mu}: ``the similarity between the
      $s\bar{s}$ spectrum, the light meson $q\bar{q}$ and the
      heavy $Q\bar{Q}$ systems needs to be understood to bridge
      the gap between Heavy Quark Effective Theory (HQET) and
      the light quark world in which we live''. }. The
      conventional {\it strangeonia} mesons are associated with
      the radial and orbital excited states of the $\phi(1020)$
      meson, the ground state of the $s\bar{s}$ system. Even
      these "normal" strangeonia are poorly understood: among
      the 22 low mass (M $<$ 2.5 GeV) strangeonium states
      expected, only 5 are well identified.
      A summary of the current data on the $\phi(1680)$, {\it i.e.} 
the first radial excitation, is shown in
      Table~\ref{strtable}. The interpretation of the current
      data is not conclusive. Photoproduction and $e^{+}e^{-}$
      annihilation experiments observed in fact different
      properties of the $\phi(1680)$ decay modes than
      hadroproduction experiments. In addition the resonance
      mass is systematically higher in photoproduction than in
      $e^{+}e^{-}$ annihilation and there is no evidence of
      $KK^{*}$ decay in photoproduction, which on the contrary
      is found to be dominant in $e^{+}e^{-}$ experiments.

      \begin{table}
      \begin{center}
      \begin{tabular}{|l|l|l|l|l|l|} \hline
      {\it  production}  & {\it mass} [MeV] & {\it width}
      (MeV) & $ {\it experiment} $ & {\it decay} & ref \\
      \hline\hline
      $e^{+}e^{-}$ & 1650 & &DM1 &
      $K_{L}K_{S}$ & ~\cite{Mane:1980ep} \\ \hline
      & 1650 & & & $K^{+}K^{-}$ & ~\cite{Delcourt:1980eq}\\ \hline
      & 1650 & & VEPP-2M & $K^{+}K^{-}$ & ~\cite{Ivanov:1981wf}\\ \hline
      & 1680 & & DM2 & $K^{+}K^{-}$ & ~\cite{Bisello:1988ez}\\ \hline
      & 1677 & 102 & & $K_{S}K^{+}\pi^{-}$.  & ~\cite{Mane:1982si}\\ \hline
      & 1680 & 185 & DM1 & $KK$, $KK\pi$ & ~\cite{Buon:1982sb}\\ \hline
      & 1657 & 146 & DM2 & $K^{+}K^{-}$ & ~\cite{Antonelli:1992jx}\\ \hline
      photo- & 1748 & 80 & CERN Omega & KK & ~\cite{Aston:1981tb} \\ \hline
      & 1760 & 80 & CERN WA57 & KK & ~\cite{Atkinson:1984cs} \\ \hline
      & 1726 & 121 & Fermi E401 & KK & ~\cite{Busenitz:1989gq} \\ \hline
      & 1753 & 122 & Fermi FOCUS& KK & ~\cite{Link:2002mp} \\ \hline
      \end{tabular}
      \caption{\small Experimental data on the $\phi(1680)$.}
      \label{strtable}
      \end{center}
      \end{table}

      The different behavior of the $\phi(1680)$ observed in
      the two types of experiment may be explained by the
      presence of two resonances interfering with the light $q\bar{q}$ 
      states. To understand this problem one could measure the
      relative branching ratios of the $\phi(1680)$ into the
      neutral and charged $KK$ and $KK^{*}$ pairs. Another
      possibility is to study the $\phi\eta$ decay mode since,
      according to the Zweig rule, the contribution of
      $s\bar{s}$ states is expected and interference with the light
      $q\bar{q}$ states should be highly suppressed. Just the
      mere identification of a resonance in the $\phi\eta$
      system will prove the presence of a $s\bar{s}$ state.
      This argument can be extended to the $\phi(1850)$ and
      other higher mass excitations.
      This decay mode has not been yet observed and is one of
      the main reactions of interest of the MesonEx program.
      Preliminary analysis of data taken with CLAS
      are promising, indicating that extraction of
      strangeonium resonances is feasible. However, due to the
      limited statistics and energy range, the current data
      will provide only a partial insight of the strangeonium
      sector. 
Higher beam energy together with an extended kaon
      identification would allow for a much
      deeper investigation of these states.
The present CLAS12
      particle identification detectors provide good $K/\pi$
      separation up to momenta of about 2.5 GeV, which corresponds to
      only a small fraction of the expected momentum range for kaons
      produced by the decay of strangeonia, which extends up to 7 GeV. 
The addition of a
      RICH detector would greatly extend the range for measuring kaons,
      resulting in a dramatic increase of acceptance for the
      investigation of these meson states
 and making the upgraded
      Hall B the ideal place to study strangeonium production.

\clearpage
\section{Acknowledgement}
We like to thank
Y. Amhis, 
A. Bacchetta,
A. Bracco, 
W. Brooks,
V. Burkert,
M. Calvetti,
E. Christova,
A. Courtoy,
L. Elouadrhiri,
R. Forty,
K. Gallmeister,
I. Garzia,
F. Giordano,
S. Gliske,
K. Griffioen,
K. Hafidi,
M. Hoek,
T. Iijima,
A. Kotzinian,
P. Krizan,
P. Kroll,
F. Kunne,
S. Liuti,
A. Martin,
H. Matevosyan,
C. Matteuzzi,
S. Melis,
F. Murgia,
C. Pauly,
J. Price,
J. Rojo Chacon,
A. Rostomyan,
S. Stepanyan,
I. Strakovsky,
O. Teryaev,
L. Trentadue,
F. Tessarotto,
and all participants of the workshop 
"Probing Strangeness in Hard Processes" (PSHP2010)
held in Frascati, Italy in October 2010,
for their contribution and very fruitful and stimulating discussions.
This workshop was supported by 
the Argonne National Laboratory,
the Laboratory Nazionali di Frascati of the Istituto Nazionale di Fisica Nucleare,
the Jefferson Science Associates, 
the Thomas Jefferson National Accelerator Facility,
and the University of Connecticut.

\newpage

\renewcommand{\baselinestretch}{1.0}
\bibliography{kaons}

\begin{thebibliography}{100}

\bibitem{PSHP-workshop}
International Workshop on Probing Strangeness in Hard Processes - PSHP2010,
  http://www.lnf.infn.it/conference/pshp2010/.

\bibitem{JLab12-TDR}
The JLab12 Technical Design Report, http://www.jlab.org/12GeV/.

\bibitem{Forte:2010dt}
S. Forte,
\newblock Acta Phys. Polon. B41 (2010) 2859, 1011.5247.

\bibitem{Lai:2007dq}
H.L. Lai et~al.,
\newblock JHEP 04 (2007) 089, hep-ph/0702268.

\bibitem{Ball:2010de}
R.D. Ball et~al.,
\newblock Nucl. Phys. B838 (2010) 136, 1002.4407.

\bibitem{Ball:2011uy}
NNPDF, R.D. Ball et~al.,
\newblock Nucl. Phys. B855 (2012) 153, 1107.2652.

\bibitem{Lai:2010vv}
H.L. Lai et~al.,
\newblock Phys. Rev. D82 (2010) 074024, 1007.2241.

\bibitem{Martin:2009iq}
A.D. Martin et~al.,
\newblock Eur. Phys. J. C63 (2009) 189, 0901.0002.

\bibitem{Airapetian:2008qf}
HERMES, A. Airapetian et~al.,
\newblock Phys. Lett. B666 (2008) 446, 0803.2993.

\bibitem{Gluck:2000dy}
M. Gluck et~al.,
\newblock Phys. Rev. D63 (2001) 094005, hep-ph/0011215.

\bibitem{deFlorian:2005mw}
D. de~Florian, G.A. Navarro and R. Sassot,
\newblock Phys. Rev. D71 (2005) 094018, hep-ph/0504155.

\bibitem{Navarro:2006bb}
G.A. Navarro and R. Sassot,
\newblock Phys. Rev. D74 (2006) 011502, hep-ph/0605266.

\bibitem{deFlorian:2008mr}
D. de~Florian et~al.,
\newblock Phys. Rev. Lett. 101 (2008) 072001, 0804.0422.

\bibitem{deFlorian:2009vb}
D. de~Florian et~al.,
\newblock Phys. Rev. D80 (2009) 034030, 0904.3821.

\bibitem{deFlorian:2007aj}
D. de~Florian, R. Sassot and M. Stratmann,
\newblock Phys. Rev. D75 (2007) 114010, hep-ph/0703242.

\bibitem{Alekseev:2010ub}
COMPASS, M.G. Alekseev et~al.,
\newblock Phys. Lett. B693 (2010) 227, 1007.4061.

\bibitem{deFlorian:2007nk}
D. de~Florian, R. Sassot and M. Stratmann,
\newblock J. Phys. Conf. Ser. 110 (2008) 022045, 0708.0769.

\bibitem{Kretzer:2000yf}
S. Kretzer,
\newblock Phys. Rev. D62 (2000) 054001, hep-ph/0003177.

\bibitem{Albino:2005me}
S. Albino, B.A. Kniehl and G. Kramer,
\newblock Nucl. Phys. B725 (2005) 181, hep-ph/0502188.

\bibitem{Mulders:1995dh}
P.J. Mulders and R.D. Tangerman,
\newblock Nucl. Phys. B461 (1996) 197, hep-ph/9510301.

\bibitem{Bacchetta:2006tn}
A. Bacchetta et~al.,
\newblock JHEP 02 (2007) 093, hep-ph/0611265.

\bibitem{Collins:1981uk}
J.C. Collins and D.E. Soper,
\newblock Nucl. Phys. B193 (1981) 381.

\bibitem{Ji:2004wu}
X. Ji, J. Ma and F. Yuan,
\newblock Phys. Rev. D71 (2005) 034005, hep-ph/0404183.

\bibitem{Sivers:1989cc}
D.W. Sivers,
\newblock Phys. Rev. D41 (1990) 83.

\bibitem{Boer:1997nt}
D. Boer and P.J. Mulders,
\newblock Phys. Rev. D57 (1998) 5780, hep-ph/9711485.

\bibitem{Collins:1992kk}
J.C. Collins,
\newblock Nucl. Phys. B396 (1993) 161, hep-ph/9208213.

\bibitem{Brodsky:2002cx}
S.J. Brodsky, D.S. Hwang and I. Schmidt,
\newblock Phys. Lett. B530 (2002) 99, hep-ph/0201296.

\bibitem{Collins:2002kn}
J.C. Collins,
\newblock Phys. Lett. B536 (2002) 43, hep-ph/0204004.

\bibitem{Belitsky:2002sm}
A.V. Belitsky, X. Ji and F. Yuan,
\newblock Nucl. Phys. B656 (2003) 165, hep-ph/0208038.

\bibitem{Airapetian:1999tv}
HERMES, A. Airapetian et~al.,
\newblock Phys. Rev. Lett. 84 (2000) 4047, hep-ex/9910062.

\bibitem{Airapetian:2001eg}
HERMES, A. Airapetian et~al.,
\newblock Phys. Rev. D64 (2001) 097101, hep-ex/0104005.

\bibitem{Airapetian:2002mf}
HERMES, A. Airapetian et~al.,
\newblock Phys. Lett. B562 (2003) 182, hep-ex/0212039.

\bibitem{Airapetian:2004tw}
HERMES, A. Airapetian et~al.,
\newblock Phys. Rev. Lett. 94 (2005) 012002, hep-ex/0408013.

\bibitem{Airapetian:2005jc}
HERMES, A. Airapetian et~al.,
\newblock Phys. Lett. B622 (2005) 14, hep-ex/0505042.

\bibitem{Airapetian:2006rx}
HERMES, A. Airapetian et~al.,
\newblock Phys. Lett. B648 (2007) 164, hep-ex/0612059.

\bibitem{Airapetian:2009ti}
HERMES, A. Airapetian et~al.,
\newblock Phys. Rev. Lett. 103 (2009) 152002, 0906.3918.

\bibitem{Airapetian:2010ds}
HERMES, A. Airapetian et~al.,
\newblock Phys. Lett. B693 (2010) 11, 1006.4221.

\bibitem{Alexakhin:2005iw}
COMPASS, V.Y. Alexakhin et~al.,
\newblock Phys. Rev. Lett. 94 (2005) 202002, hep-ex/0503002.

\bibitem{Alekseev:2008dn}
COMPASS, M. Alekseev et~al.,
\newblock Phys. Lett. B673 (2009) 127, 0802.2160.

\bibitem{Alekseev:2010rw}
COMPASS, M.G. Alekseev et~al.,
\newblock Phys. Lett. B692 (2010) 240, 1005.5609.

\bibitem{Bradamante:2011xu}
COMPASS, F. Bradamante,
\newblock (2011), 1111.0869.

\bibitem{Avakian:2003pk}
CLAS, H. Avakian et~al.,
\newblock Phys. Rev. D69 (2004) 112004, hep-ex/0301005.

\bibitem{Avakian:2005ps}
CLAS, H. Avakian et~al.,
\newblock AIP Conf. Proc. 792 (2005) 945, nucl-ex/0509032.

\bibitem{Osipenko:2008rv}
CLAS, M. Osipenko et~al.,
\newblock Phys. Rev. D80 (2009) 032004, 0809.1153.

\bibitem{Avakian:2010ae}
CLAS, H. Avakian et~al.,
\newblock Phys. Rev. Lett. 105 (2010) 262002, 1003.4549.

\bibitem{Aghasyan:2011ha}
CLAS, M. Aghasyan et~al.,
\newblock Phys.Lett. B704 (2011) 397, 1106.2293.

\bibitem{Lee:2009ck}
BRAHMS, J.H. Lee and F. Videbaek,
\newblock (2009), 0908.4551.

\bibitem{Arsene:2008mi}
BRAHMS, I. Arsene et~al.,
\newblock Phys. Rev. Lett. 101 (2008) 042001, 0801.1078.

\bibitem{Adler:2005in}
PHENIX, S.S. Adler et~al.,
\newblock Phys. Rev. Lett. 95 (2005) 202001, hep-ex/0507073.

\bibitem{Chiu:2007zy}
PHENIX, M. Chiu,
\newblock AIP Conf. Proc. 915 (2007) 539, nucl-ex/0701031.

\bibitem{Adare:2010bd}
PHENIX, A. Adare et~al.,
\newblock Phys. Rev. D82 (2010) 112008, 1009.4864.

\bibitem{Dharmawardane:2010zz}
PHENIX, V. Dharmawardane,
\newblock PoS DIS2010 (2010) 222.

\bibitem{Adams:2003fx}
STAR, J. Adams et~al.,
\newblock Phys. Rev. Lett. 92 (2004) 171801, hep-ex/0310058.

\bibitem{Abelev:2007ii}
STAR, B.I. Abelev et~al.,
\newblock Phys. Rev. Lett. 99 (2007) 142003, 0705.4629.

\bibitem{Abelev:2008qb}
STAR, B.I. Abelev et~al.,
\newblock Phys. Rev. Lett. 101 (2008) 222001, 0801.2990.

\bibitem{Eun:2010zz}
STAR, L.K. Eun,
\newblock J. Phys. Conf. Ser. 230 (2010) 012041.

\bibitem{Abe:2005zx}
BELLE, K. Abe et~al.,
\newblock Phys. Rev. Lett. 96 (2006) 232002, hep-ex/0507063.

\bibitem{Seidl:2008xc}
BELLE, R. Seidl et~al.,
\newblock Phys. Rev. D78 (2008) 032011, 0805.2975.

\bibitem{Garzia:2011}
BABAR, I. Garzia,
\newblock (2012), 1201.4678,
\newblock Proceedings of Transversity 2011.

\bibitem{Kotzinian:1994dv}
A. Kotzinian,
\newblock Nucl. Phys. B441 (1995) 234, hep-ph/9412283.

\bibitem{Diehl:2005pc}
M. Diehl and S. Sapeta,
\newblock Eur. Phys. J. C41 (2005) 515, hep-ph/0503023.

\bibitem{Giordano:2010gq}
HERMES, F. Giordano and R. Lamb,
\newblock J.Phys.Conf.Ser. 295 (2011) 012092, 1011.5422.

\bibitem{Cahn:1978se}
R.N. Cahn,
\newblock Phys. Lett. B78 (1978) 269.

\bibitem{Falciano:1986wk}
NA10, S. Falciano et~al.,
\newblock Z. Phys. C31 (1986) 513.

\bibitem{Guanziroli:1987rp}
NA10, M. Guanziroli et~al.,
\newblock Z. Phys. C37 (1988) 545.

\bibitem{Zhu:2006gx}
FNAL-E866/NuSea, L.Y. Zhu et~al.,
\newblock Phys. Rev. Lett. 99 (2007) 082301, hep-ex/0609005.

\bibitem{Lam:1978zr}
C.S. Lam and W.K. Tung,
\newblock Phys. Lett. B80 (1979) 228.

\bibitem{Boer:1999si}
D. Boer and P.J. Mulders,
\newblock Nucl. Phys. B569 (2000) 505, hep-ph/9906223.

\bibitem{Boer:1997mf}
D. Boer, R. Jakob and P.J. Mulders,
\newblock Nucl. Phys. B504 (1997) 345, hep-ph/9702281.

\bibitem{Anselmino:2007fs}
M. Anselmino et~al.,
\newblock Phys.Rev. D75 (2007) 054032, hep-ph/0701006.

\bibitem{Anselmino:2008jk}
M. Anselmino et~al.,
\newblock Nucl. Phys. Proc. Suppl. 191 (2009) 98, 0812.4366.

\bibitem{Bacchetta:2007wc}
A. Bacchetta et~al.,
\newblock Phys. Lett. B659 (2008) 234, 0707.3372.

\bibitem{Collins:1993kq}
J.C. Collins, S.F. Heppelmann and G.A. Ladinsky,
\newblock Nucl. Phys. B420 (1994) 565, hep-ph/9305309.

\bibitem{Bianconi:1999cd}
A. Bianconi et~al.,
\newblock Phys. Rev. D62 (2000) 034008, hep-ph/9907475.

\bibitem{Vossen:2011fk}
BELLE, A. Vossen et~al.,
\newblock Phys.Rev.Lett. 107 (2011) 072004, 1104.2425.

\bibitem{Bacchetta:2003vn}
A. Bacchetta and M. Radici,
\newblock Phys. Rev. D69 (2004) 074026, hep-ph/0311173.

\bibitem{Airapetian:2008sk}
HERMES, A. Airapetian et~al.,
\newblock JHEP 06 (2008) 017, 0803.2367.

\bibitem{Gliske-phd:2011}
S. Gliske,
\newblock Ph.D. thesis, University of Michigan  (2011).

\bibitem{Martin:2007au}
COMPASS, A. Martin,
\newblock Czech. J. Phys. 56 (2006) F33, hep-ex/0702002.

\bibitem{Wollny:2009eq}
COMPASS, H. Wollny,
\newblock (2009), 0907.0961.

\bibitem{Braun:2012}
COMPASS, C. Braun,
\newblock To appear in the Transversity 2011 proceedings  (2012), 

\bibitem{Courtoy:2011ni}
A. Courtoy, A. Bacchetta and M. Radici,
\newblock (2011), 1106.5897.

\bibitem{Kane:1978nd}
G.L. Kane, J. Pumplin and W. Repko,
\newblock Phys. Rev. Lett. 41 (1978) 1689.

\bibitem{Adams:1991rx}
FNAL-E581, D.L. Adams et~al.,
\newblock Phys. Lett. B261 (1991) 197.

\bibitem{Adams:1991ru}
FNAL-E581, D.L. Adams et~al.,
\newblock Z. Phys. C56 (1992) 181.

\bibitem{Adams:1991cs}
FNAL-E704, D.L. Adams et~al.,
\newblock Phys. Lett. B264 (1991) 462.

\bibitem{Adams:1994yu}
FNAL-E704, D.L. Adams et~al.,
\newblock Phys. Rev. D53 (1996) 4747.

\bibitem{Bravar:1996ki}
FNAL-E704, A. Bravar et~al.,
\newblock Phys. Rev. Lett. 77 (1996) 2626.

\bibitem{D'Alesio:2007jt}
U. D'Alesio and F. Murgia,
\newblock Prog. Part. Nucl. Phys. 61 (2008) 394, 0712.4328.

\bibitem{Mueller:1998fv}
D. Mueller et~al.,
\newblock Fortsch.Phys. 42 (1994) 101, hep-ph/9812448.

\bibitem{Radyushkin:1996nd}
A. Radyushkin,
\newblock Phys.Lett. B380 (1996) 417, hep-ph/9604317.

\bibitem{Ji:1996nm}
X.D. Ji,
\newblock Phys.Rev. D55 (1997) 7114, hep-ph/9609381.

\bibitem{Ji:1996ek}
X.D. Ji,
\newblock Phys.Rev.Lett. 78 (1997) 610, hep-ph/9603249.

\bibitem{Goloskokov:2005sd}
S. Goloskokov and P. Kroll,
\newblock Eur.Phys.J. C42 (2005) 281, hep-ph/0501242.

\bibitem{Goloskokov:2007nt}
S. Goloskokov and P. Kroll,
\newblock Eur.Phys.J. C53 (2008) 367, 0708.3569.

\bibitem{Goloskokov:2008ib}
S. Goloskokov and P. Kroll,
\newblock Eur.Phys.J. C59 (2009) 809, 0809.4126.

\bibitem{Goloskokov:2011rd}
S. Goloskokov and P. Kroll,
\newblock Eur.Phys.J. A47 (2011) 112, 1106.4897.

\bibitem{Airapetian:2011jp}
HERMES, A. Airapetian et~al.,
\newblock Eur. Phys. J. A47 (2011) 113, 1107.3496.

\bibitem{Airapetian:2009jy}
HERMES Collaboration, A. Airapetian et~al.,
\newblock Phys.Lett. B684 (2010) 114, 0906.2478.

\bibitem{Gao:2010mj}
J.H. Gao, Z.t. Liang and X.N. Wang,
\newblock Phys.Rev. C81 (2010) 065211, 1001.3146.

\bibitem{Nakamura:2010zzi}
Particle Data Group, K. Nakamura et~al.,
\newblock J. Phys. G37 (2010) 075021.

\bibitem{Jaffe:1975fd}
R.L. Jaffe and K. Johnson,
\newblock Phys. Lett. B60 (1976) 201.

\bibitem{Horn:1977rq}
D. Horn and J. Mandula,
\newblock Phys. Rev. D17 (1978) 898.

\bibitem{Barnes:1982tx}
T. Barnes et~al.,
\newblock Nucl. Phys. B224 (1983) 241.

\bibitem{Isgur:1985vy}
N. Isgur, R. Kokoski and J.E. Paton,
\newblock Phys. Rev. Lett. 54 (1985) 869.

\bibitem{Isgur:1984bm}
N. Isgur and J.E. Paton,
\newblock Phys. Rev. D31 (1985) 2910.

\bibitem{Guo:2008yz}
P. Guo et~al.,
\newblock Phys. Rev. D78 (2008) 056003, 0807.2721.

\bibitem{Dudek:2009qf}
J.J. Dudek et~al.,
\newblock Phys. Rev. Lett. 103 (2009) 262001, 0909.0200.

\bibitem{Dudek:2010wm}
J.J. Dudek et~al.,
\newblock Phys. Rev. D82 (2010) 034508, 1004.4930.

\bibitem{Close:1978be}
F.E. Close and H.J. Lipkin,
\newblock Phys. Rev. Lett. 41 (1978) 1263.

\bibitem{Bityukov:2005ag}
S.I. Bityukov et~al.,
\newblock (2005), physics/0512056.

\bibitem{Bityukov:1986yd}
S.I. Bityukov et~al.,
\newblock Phys. Lett. 188B (1987) 383.

\bibitem{Atkinson:1983df}
OMEGA PHOTON, M. Atkinson et~al.,
\newblock Nucl. Phys. B231 (1984) 1.

\bibitem{Antipov:1995kb}
Y.M. Antipov et~al.,
\newblock Instrum. Exp. Tech. 38 (1995) 581.

\bibitem{Aubert:2008jd}
BABAR, B. Aubert et~al.,
\newblock Phys. Rev. D78 (2008) 092002, 0807.2408.

\bibitem{Barnes:2002mu}
T. Barnes, N. Black and P.R. Page,
\newblock Phys. Rev. D68 (2003) 054014, nucl-th/0208072.

\bibitem{Mane:1980ep}
F. Mane et~al.,
\newblock Phys. Lett. B99 (1981) 261.

\bibitem{Delcourt:1980eq}
B. Delcourt et~al.,
\newblock Phys. Lett. B99 (1981) 257.

\bibitem{Ivanov:1981wf}
P.M. Ivanov et~al.,
\newblock Phys. Lett. B107 (1981) 297.

\bibitem{Bisello:1988ez}
DM2, D. Bisello et~al.,
\newblock Z. Phys. C39 (1988) 13.

\bibitem{Mane:1982si}
F. Mane et~al.,
\newblock Phys. Lett. B112 (1982) 178.

\bibitem{Buon:1982sb}
J. Buon et~al.,
\newblock Phys. Lett. B118 (1982) 221.

\bibitem{Antonelli:1992jx}
DM2, A. Antonelli et~al.,
\newblock Z.Phys. C56 (1992) 15.

\bibitem{Aston:1981tb}
D. Aston et~al.,
\newblock Phys. Lett. B104 (1981) 231.

\bibitem{Atkinson:1984cs}
OMEGA PHOTON, M. Atkinson et~al.,
\newblock Z. Phys. C27 (1985) 233.

\bibitem{Busenitz:1989gq}
J. Busenitz et~al.,
\newblock Phys. Rev. D40 (1989) 1.

\bibitem{Link:2002mp}
FOCUS, J.M. Link et~al.,
\newblock Phys. Lett. B545 (2002) 50, hep-ex/0208027.

\end{thebibliography}






\end{document}